\documentclass[journal,twoside,web]{IEEEtran}
\usepackage{tmi}
\usepackage{cite}
\usepackage{amsmath,amssymb,amsfonts}
\usepackage{siunitx}
\usepackage{mathtools}
\usepackage{algorithm}

\newcommand{\revend}{\normalcolor}
\usepackage{diagbox}
\usepackage{algpseudocode}
\usepackage[pdftex]{graphicx}
\usepackage{bm}
\usepackage{hyperref}

\def\E{\mathbb{E}}

\def\xbm{{\bm{x}}}
\def\zbm{{\bm{z}}}

\def\zbm{{\bm{z}}}

\def\thetabm{{\bm{\theta }}}

\def\Ncal{{\mathcal{N}}}

\usepackage[switch]{lineno}

\newcommand{\norm}[1]{\left\lVert#1\right\rVert}

\UseRawInputEncoding
\def\BibTeX{{\rm B\kern-.05em{\sc i\kern-.025em b}\kern-.08em
    T\kern-.1667em\lower.7ex\hbox{E}\kern-.125emX}}
\begin{document}
\title{Dose-aware Diffusion Model for 3D PET Image Denoising: Multi-institutional Validation with Reader Study and Real Low-dose Data}
\author{
Huidong Xie$^{1}$ \IEEEmembership{Student Member, IEEE}, Weijie Gan$^2$ \IEEEmembership{Student Member, IEEE}, Reimund Bayerlein$^3$, Bo Zhou$^{1}$, Ming-Kai Chen$^{4}$, Michal Kulon$^4$, Annemarie Boustani$^4$, Kuan-Yin Ko$^5$, Der-Shiun Wang$^{6,7}$, Benjamin A. Spencer$^3$, Wei Ji$^{4}$, Xiongchao Chen$^{1}$, Qiong Liu$^{1}$,  Xueqi Guo$^{1}$, Menghua Xia$^{4}$, Yinchi Zhou$^{1}$, \\ Hui Liu$^{4}$, Liang Guo$^{1}$, Hongyu An$^8$, Ulugbek S. Kamilov$^{2,9}$, Hanzhong Wang$^{10}$, Biao Li$^{10}$, Axel Rominger$^{11}$, \\Kuangyu Shi$^{11}$, Ge Wang$^{12}$ \IEEEmembership{Life Fellow, IEEE}, Ramsey D. Badawi$^{3,13}$ \IEEEmembership{Senior Member, IEEE}, Chi Liu$^{1,4}$\IEEEmembership{Senior Member, IEEE}.
\thanks{Corresponding author: Chi Liu.}
\thanks{Emails: \{Huidong.Xie; Chi.Liu\}@yale.edu}
\thanks{$^1$Department of Biomedical Engineering, Yale University, USA.}
\thanks{$^2$Department of Computer Science, Washington University in St. Louis, USA.}
\thanks{$^3$Department of Radiology, University of California Davis, USA.}
\thanks{$^4$Department of Radiology and Biomedical Imaging, Yale University, USA.}
\thanks{$^5$Department of Nuclear Medicine, National Taiwan University Cancer Center, Taipei, Taiwan}
\thanks{$^6$Department of Pediatric, Tri-service General Hospital, Taipei, Taiwan.}
\thanks{$^7$Department of Pediatric, School of Medicine, National Defenese Medical Center, Taipei, Taiwan.}
\thanks{$^8$Department of Radiology, Washington University in St. Louis, USA.}
\thanks{$^9$Department of Electrical \& Systems Engineering, Washington University in St. Louis, USA.}
\thanks{$^{10}$Department of Nuclear Medicine, Ruijin Hospital, Shanghai Jiao Tong University School of Medicine, China.}
\thanks{$^{11}$Department of Nuclear Medicine, University of Bern, Switzerland.}
\thanks{$^{12}$Department of Biomedical Engineering, Rensselaer Polytechnic Institute, USA.}
\thanks{$^{13}$Department of Biomedical Engineering, University of California Davis, USA.}}
\maketitle

\begin{abstract}
Reducing scan times, radiation dose, and enhancing image quality, especially for lower-performance scanners, are critical in low-count/low-dose PET imaging. Deep learning (DL) techniques have been investigated for PET image denoising. However, existing models have often resulted in compromised image quality when achieving low-count/low-dose PET and have limited generalizability to different image noise-levels, acquisition protocols, and patient populations. Recently, diffusion models have emerged as the new state-of-the-art generative model to generate high-quality samples and have demonstrated strong potential for medical imaging tasks. However, for low-dose PET imaging, existing diffusion models failed to generate consistent 3D reconstructions, unable to generalize across varying noise-levels, often produced visually-appealing but distorted image details, and produced images with biased tracer uptake. Here, we develop DDPET-3D, a dose-aware diffusion model for 3D low-dose PET imaging to address these challenges. Collected from 4 medical centers globally with different scanners and clinical protocols, we extensively evaluated the proposed model using a total of 9,783 $^{18}\text{F-FDG}$ studies (1,596 patients) with low-dose/low-count levels ranging from 1\% to 50\%. With a cross-center, cross-scanner validation, the proposed DDPET-3D demonstrated its potential to generalize to different low-dose levels, different scanners, and different clinical protocols. As confirmed with reader studies performed by board-certified nuclear medicine physicians, experienced readers judged the images to be similar to or superior to the full-dose images and previous DL baselines based on qualitative visual impression. We also evaluated the lesion-level quantitative accuracy using a Monte Carlo simulation study and a lesion segmentation network. The presented results show the potential to achieve low-dose PET while maintaining image quality. Lastly, a group of real low-dose scans was also included for evaluation to demonstrate the clinical potential of DDPET-3D. The code is made publicly available on GitHub: \href{https://github.com/HuidongXie/DDPET-3D}{https://github.com/HuidongXie/DDPET-3D}

\end{abstract}


\section{Introduction}
\label{sec:introduction}
Positron Emission Tomography (PET) is a functional imaging modality widely used in oncology, cardiology, and neurology studies \cite{rohren2004clinical,schwaiger2005pet,clark2012cerebral}. Given the growing concern of radiation exposure and potentially increased cancer risks associated with PET scans, reducing the PET injection dose is desirable \cite{robbins2008radiation}. In addition, reducing patient motion artifacts and patient discomfort through shorter scans as well as enhancing image quality on lower-performance scanners would also be beneficial in clinical settings. However, reduced injection dose and shorter scan duration can negatively affect PET image quality, and may affect diagnostic performance such as the identification of low-contrast lesions \cite{schaefferkoetter2015initial}. Therefore, reconstructing high-quality images from noisy input is an important topic. In fact, making the radiation dose as low as reasonably achievable (the ALARA principle) is a commonly accepted practice in clinical settings.


Many different deep learning (DL) methods were proposed for low-dose/low-count PET image reconstructions/restorations \cite{xu_200x_2017, zhou2020supervised, ouyang_ultra-low-dose_2019, zhou2021mdpet, zhou2023fast, zhou_federated_2023, zhou2023fedftn, liu2022personalized, gong_pet_2019, onishi_anatomical-guided_2021, xie_unified_2023, liu_pet_2022}. In particular, with the advancement of cutting-edge generative models \cite{Croitoru.etal2023,Kazerouni.etal2023a}, diffusion models have demonstrated strong potential for low-dose PET imaging \cite{gong_pet_2023, jiang_pet-diffusion_2023}. Diffusion models have also been proposed for other imaging modalities, such as CT \cite{gao_corediff_2023} and MRI \cite{gungor_adaptive_2023, chung_score-based_2022}. However, we observe that most of these methods focus on 2D imaging, leaving the application of diffusion models for 3D PET imaging largely underexplored in the literature. Several challenges should be addressed to effectively apply diffusion models to 3D PET imaging.

\textbf{First}, PET imaging (PET) is intrinsically a 3D imaging modality, whereas previous 2D diffusion works do not take into account the 3D spatial correlation between different slices. Thus, a 3D diffusion model is desirable. However, due to the hardware memory limit, directly extending the diffusion model to 3D would be difficult. 

\textbf{Second}, another challenge with PET image denoising is the high variation of image noise levels among patients, especially when dealing with multi-center, cross-scanner dataset.  The noise level in PET images can be affected by various factors, including, but not limited to: \emph{(1)} Variations of post tracer injection acquisition start time. \emph{(2)} Occasional tracer injection infiltration in some patients, which causes the tracer to accumulate at the injection location, resulting in high image noise in the body. \emph{(3)} The variation of patient's BMI and associated weight-based tracer injection protocols. \emph{(4)} Different hospitals using scanners with varying sensitivities have different total scan times for patients. \emph{(5)} Variations of excretion into the urinary bladder. Higher excretion would also result in higher noise in other parts of the image. All of these factors result in variability of the noise levels in the reconstructed images. Therefore, a method to adaptively denoise images with varying noise levels is desirable. However, previously proposed methods mentioned above have limited generalizability to different noise levels. A network trained on one noise level often fails to produce high-quality reconstructions on other noise levels.

\textbf{Third}, different from other imaging modalities, PET was developed as a quantitative tool, being recognized as providing an objective, and more accurate measure for prognosis and response monitoring purposes than visual inspection alone \cite{boellaard_standards_2009}. However, our experimental results showed that although standard diffusion models (DDPM \cite{ho2020denoising}, DDIM\cite{song_denoising_2022}) produce visually appealing images, they typically fail to maintain accurate quantification and produced images with biased tracer uptake. Specifically, the tracer distribution in the entire image changes after diffusion models. This issue is more severe in whole-body PET scans since the tracer uptake could vary significantly among organs. 

In this work, we develop a dose-aware diffusion model for 3D PET Imaging (DDPET-3D) to address these limitations. The main methodological contributions of the proposed DDPET-3D framework are: \textbf{First:} We proposed a 2.5D diffusion strategy with multiple fixed noise variables to address the 3D issue in diffusion models for medical imaging. DDPET-3D maintains a similar memory burden to 2D diffusion models while achieving high-quality reconstructions. \textbf{Second:} We proposed a dose-embedding strategy that allows noise-aware denoising. DDPET-3D can simultaneously denoise 3D PET images with varying low-dose/count levels. \textbf{Third:} To achieve a quantitative-accurate reconstruction, we proposed to use a denoised prior in DDPET-3D. The proposed denoised prior also allows the diffusion model to converge within 25 sampling steps. DDPET-3D can denoise 3D PET images within a reasonable time constraint  ($\sim$15 mins on a single NVIDIA A40 GPU). Previous diffusion methods using DDPM sampling would require approximately $6$ hr. 

\textbf{Furthermore}, existing models are often tailored to specific hospitals, scanners, patient populations, acquisition protocols, and noise levels, limiting their generalizability across the diverse PET image acquisition settings encountered in practice. To address this, we evaluated the proposed method on a large-scale, cross-center, cross-scanner dataset, which is potentially the most diverse and extensive dataset in PET-related AI literature. In this work, we coordinated different hospitals around the world, prepared this large-scale dataset for the evaluation of PET denoising algorithms, and have made parts of the data publicly available. In addition to traditional image quality metrics (SSIM, PSNR, etc.), we designed a reader study with nuclear medicine physicians to provide a clinical insight of the model performance. This evaluation highlights the method's clinical potential and generalizability across various acquisition settings, making it more applicable to real-world scenarios.

\begin{figure*}[!t]
\centerline{\includegraphics[width=\textwidth]{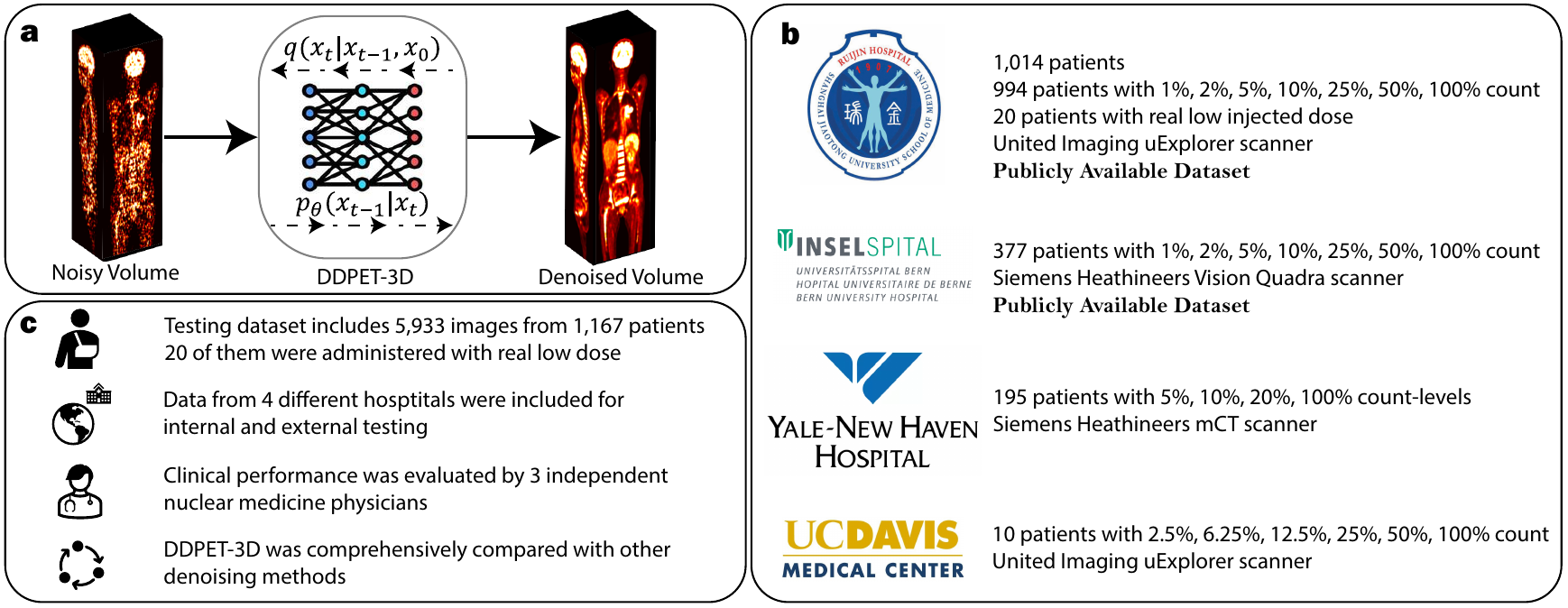}}
\caption{General overview of the study. a) Model development. The proposed diffusion network, DDPET-3D, was trained in a supervised manner with paired low-count/full-count images. DDPET-3D takes low-count/low-dose 3D volumes as input and outputs the synthesized the corresponding full-count/normal-dose images. Further details on network structure are provided later in this paper (see Methods). b), Dataset overview. 1,596 patient studies obtained at 4 different hospitals with different scanners, clinical protocols, and low-count/low-dose levels were included in this study. c), Model evaluations: 1,167 patients from 4 hospitals were included for internal and external model evaluations. Reader studies were performed to assess the image quality. DDPET-3D was also compared with other denoising methods.}
\label{overview_fig}
\end{figure*}

The ultimate goal of DL methods for PET image denoising is to achieve low-dose PET without compromising image quality and clinical performance. As outlined in Fig. \ref{overview_fig}(c), evaluated on real-world, large-scale, cross-center, ultra-low-dose 3D PET data, with confirmation from reader studies with nuclear medicine physicians, DDPET-3D demonstrates superior quantitative and qualitative results that are comparable or even better than the 100\% full-count data as well as previous baseline methods across varying low-count-levels. DDPET-3D also showed superior generalizability when directly applied to data acquired at a different hospital without further network fine-tuning. We also evaluated the proposed method on a cohort of human studies with real low-dose injection. The results presented show the potential of DDPET-3D to achieve low-dose PET imaging while maintaining image quality. An overview of the dataset is provided in Fig. \ref{overview_fig}(b).
\section{Related Work}

Deep learning methods for PET image denoising can be classified into three broad categories: (1) convolutional-based models, (2) transformer-based models, and (3) diffusion-based models.

\textbf{Convolutional-based models} were among the first approaches to low-dose PET denoising. Many supervised networks were proposed to learn a mapping between noisy low-quality and high-quality images, including U-net variants \cite{xu_200x_2017, liu2022personalized, liu_pet_2022, mehranian_image_2022, schaefferkoetter_convolutional_2020}, residual learning \cite{gong_pet_2019}, and generative adversarial networks (GANs) \cite{ouyang_ultra-low-dose_2019, zhou2021mdpet, wang_3d_2018}. While these methods show promising visual performance, they typically fail to generalize across different acquisition settings (e.g., different low-dose levels). To address this, we previously proposed a Unified Noise-Aware Network (UNN) \cite{xie_unified_2023} to integrate multiple U-net \cite{ronneberger_u-net_2015} sub-networks with varying denoising power to achieve adaptive and noise-aware denoising. However, training multiple sub-networks requires tedious data preprocessing and a long training time. The testing time also linearly increases with the number of sub-networks. Unsupervised and self-supervised methods have also been explored in the literature \cite{xie_noise-aware_2025, song_noise2void_2021, cui_pet_2019, onishi_anatomical-guided_2021}, but they are beyond the scope of this paper.

\textbf{Transformer-based models} have been explored in PET image denoising mainly due to their ability to model long-range dependencies and capture global contextual information. For example, Jang \textit{ et al.} presented Spach Transformer \cite{jang2023spach}, an efficient spatial and channel-wise encoder decoder transformer that can leverage spatial and channel information based on local and global global multi-head self-attention. Yang  \textit{ et al.} introduced AMIR (All-In-One Medical Image Restoration) network \cite{yang2024amir}, a transformer-based architecture to address different medical imaging problems using one unified model.

\textbf{Diffusion-based models} have demonstrated strong potential for PET image denoising by progressively refining noisy inputs through iterative denoising steps rooted in probabilistic modeling. For example, based on the DDPM framework \cite{ho2020denoising}, Gong \textit{et al.} proposed to perform PET image denoising with MRI as prior information to improve image quality \cite{gong_pet_2023}. Unfortunately, PET-MR systems are not widely available and it would be time consuming to obtain sequential and paired PET-MR images in reality for network training and image reconstructions. Jiang \textit{et al.}\cite{jiang_pet-diffusion_2023} adopted the latent diffusion model \cite{rombach_high-resolution_2022} for unsupervised PET denoising.

As mentioned above, directly training a 3D diffusion model for 3D imaging is difficult due to hardware limit. Several previous works aim to address the 3D imaging problems of diffusion models. For example, Chung \textit{et al.}\cite{chung_solving_2023} proposed to apply a TV penalty term along the z-axis to remove inconsistencies within each reverse sampling step in the diffusion model. On the other hand, Lee \textit{et al.}\cite{lee_improving_2023} used two pretrained perpendicular 2D diffusion models to remove inconsistencies between slices. However, presented later in this paper, these previous methods did not produce satisfactory results for 3D low-count PET (e.g., inaccurate organ boundaries in extreme low-dose settings, inconsistent 3D reconstructions, and distorted image features in some cases). Yoon \textit{et al.} presented Conditional Score-based Residual Diffusion (CSRD) \cite{yoon_volumetric_2024} which adapted a patch-based training strategy to alleviate the memory bottleneck associated with the 3D diffusion model. While CSRD provides a feasible path toward a volumetric diffusion model, due to the high memory demands of pure 3D diffusion models, the largest path size we could process into the GPU was smaller than the 2.5D conditional volume used in the proposed DDPET-3D method, potentially limiting its effectiveness on whole-body PET data.

In this paper, representative methods in each category were selected and compared.
\section{Methods}
\subsection{Preliminary: Diffusion models}
The general idea of diffusion models is to learn the target data distribution $q(\xbm_0)$ (\emph{i.e.,} full-dose PET images in our case) using neural networks. Once the distribution is learned, we can synthesize a new sample from it. Diffusion models consist of two Markov chains: the forward diffusion process and the learned reverse diffusion process.
The forward diffusion process $q$ gradually adds a small amount of Gaussian noise to $\xbm_0\sim q(\xbm_0)$ at each step, until the original image signal is completely destroyed. As defined in \cite{ho2020denoising} 
\begin{equation}
\begin{split}
q(\xbm_{1:T}|\xbm_0)&\coloneqq \prod_{t=1}^Tq(\xbm_t|\xbm_{t-1})\ , \\ \text{where} \; q(\xbm_t|\xbm_{t-1})&\coloneqq\mathcal{N}(\xbm_t;\sqrt{1-\beta_t}\xbm_{t-1},\beta_t\mathbf{I})\ .
\label{eq1}
\end{split}
\end{equation}
One property of the diffusion process is that, one can sample $\xbm_t$ for any arbitrary time-step $t$ without gradually adding noise to $\xbm_0$. By denoting $\alpha_t \coloneqq 1-\beta_t$ and $\bar{\alpha}_t \coloneqq \prod_{s=1}^t \alpha_s$, we have
\begin{equation}
\begin{split}
q(\xbm_t|\xbm_0)&=\mathcal{N}(\xbm_t;\sqrt{(\bar{\alpha}_t})\xbm_0,(1-\bar{\alpha}_t)\mathbf{I})\ , \\ 
\text{and}\ \xbm_t & = \sqrt{\bar{\alpha}_t}\xbm_0+\sqrt{1-\bar{\alpha}_t}{\bm \epsilon}\ ,
\label{eq3}
\end{split}
\end{equation}
where ${\bm \epsilon} \sim \mathcal{N}(0,\mathbf{I})$. The latent $\xbm_T$ is nearly an isotropic Gaussian distribution for a properly designed $\beta_t$ schedule. Therefore, one can easily generate a new $\xbm_T$ and then synthesize a $\xbm_0$ by progressively sampling from the reverse posterior $q(\xbm_{t-1}|\xbm_t)$.
However, this reverse posterior is tractable only if $\xbm_0$ is known
\begin{equation}
\label{equ:q-posi}
    q(\xbm_{t-1}|\xbm_t, \xbm_0) = \Ncal\Big(\xbm_{t-1}; \mu_q(\xbm_t,\xbm_0), \frac{\beta_t(1-\bar{\alpha}_{t-1})}{1-\bar{\alpha}_t}\textbf{I}\Big)\ ,
\end{equation}
where
\begin{equation}
    \mu_q(\xbm_t,\xbm_0) = \frac{\sqrt{\alpha_t}(1-\bar{\alpha}_{t-1})\xbm_t + \sqrt{\bar{\alpha}_{t-1}}(1-\alpha_t)\xbm_0}{1-\bar{\alpha}_t}\ .
\end{equation}
Note that $q(\xbm_{t-1}|\xbm_t):=q(\xbm_{t-1}|\xbm_t, \xbm_0)$, where the extra conditioning term $\xbm_0$ is superfluous due to the Markov property.
DDPM thus proposes to learn a parameterized Gaussian transitions $p_\thetabm(\xbm_{t-1}|\xbm_t)$ to approximate the reverse diffusion posterior \eqref{equ:q-posi}
\begin{equation}
\label{equ:p-posi}
    p_\thetabm(\xbm_{t-1}|\xbm_t) = \Ncal\Big(\xbm_{t-1}; \mu_\thetabm(\xbm_t,t), \sigma^2_t\textbf{I}\Big)\ ,
\end{equation}
where
\begin{equation}
    \label{equ:p-mean}
    \mu_\thetabm(\xbm_t,t) = \frac{1}{\sqrt{\alpha_t}}\Big(\xbm_t-\frac{1-\alpha_t}{\sqrt{1-\bar{\alpha}_t}}\epsilon_\thetabm(\xbm_t, t)\Big)\ .
\end{equation}
Here, $\epsilon_\thetabm$ denotes a neural network. 
Through some derivations detailed in \cite{ho2020denoising}, the training objective of $\epsilon_\thetabm(\xbm_t, t)$ can be formulated as 
\begin{equation}
    \label{equ:training}
    \E_{\xbm,{\bm\epsilon},t\sim[1,T]}\big[\norm{{\bm \epsilon} - {\epsilon}_\thetabm(\xbm_t, t)}^2\big]\ .
\end{equation}
It is worth noting that the original DDPM \cite{ho2020denoising}  set $\sigma_t$ to a fixed constant value based on the $\beta_t$ schedule. Recent studies~\cite{dhariwal2021diffusion,nichol2021improved} have shown the improved performance by using the learned variance $\sigma_t^2\coloneqq\sigma_\thetabm^2(\xbm_t, t)$. We also adopted this approach. To be specific, we have $\sigma_\thetabm(\xbm_t, t)\coloneqq\exp(v\log \beta_t + (1-v)\log\tilde{\beta}_t)$, where $\tilde{\beta}_t$ refers to the lower bound for the reverse diffusion posterior variances \cite{ho2020denoising}, and $v$ denotes the network output. We used a single neural network with two separate output channels to estimate the mean and the variance of \eqref{equ:p-mean} jointly.
Based on the learned reverse posterior $p_\thetabm(\xbm_{t-1}|\xbm_t)$, the iteration of obtaining a $\xbm_0$ from a $\xbm_T$ can be formulated as follow
\begin{equation}
    \xbm_{t-1} = \mu_\thetabm(\xbm_t,t) + \sigma_t\zbm,\text{ where } \zbm\sim\Ncal(0,\textbf{I})\ .
\end{equation}

\subsection{Conditional PET image denoising}

The framework described above only allows unconditional sampling. For the purpose of low-dose PET denoising, instead of generating new samples, the network needs to denoise the images based on input noisy counterparts. This can be achieved by adding an additional condition to the neural network. $\epsilon_\thetabm{(\xbm_t,t)}$ becomes $\epsilon_\thetabm{(\xbm_t,t,\xbm_\mathrm{noisy})}$, where $\xbm_\mathrm{noisy}$ denotes the noisy input PET images. Specifically, the input to the neural network becomes a two-channel input, one is $\xbm_t$, the other one is $\xbm_\mathrm{noisy}$. However, we noticed that this technique results in severe inconsistencies in the reconstructed 3D image volumes. One example is presented in Fig. \ref{UI_results_compare} (DDIM results).

\subsection{Proposed DDPET-3D}

\begin{figure*}[!ht]
\centerline{\includegraphics[width=0.8\textwidth]{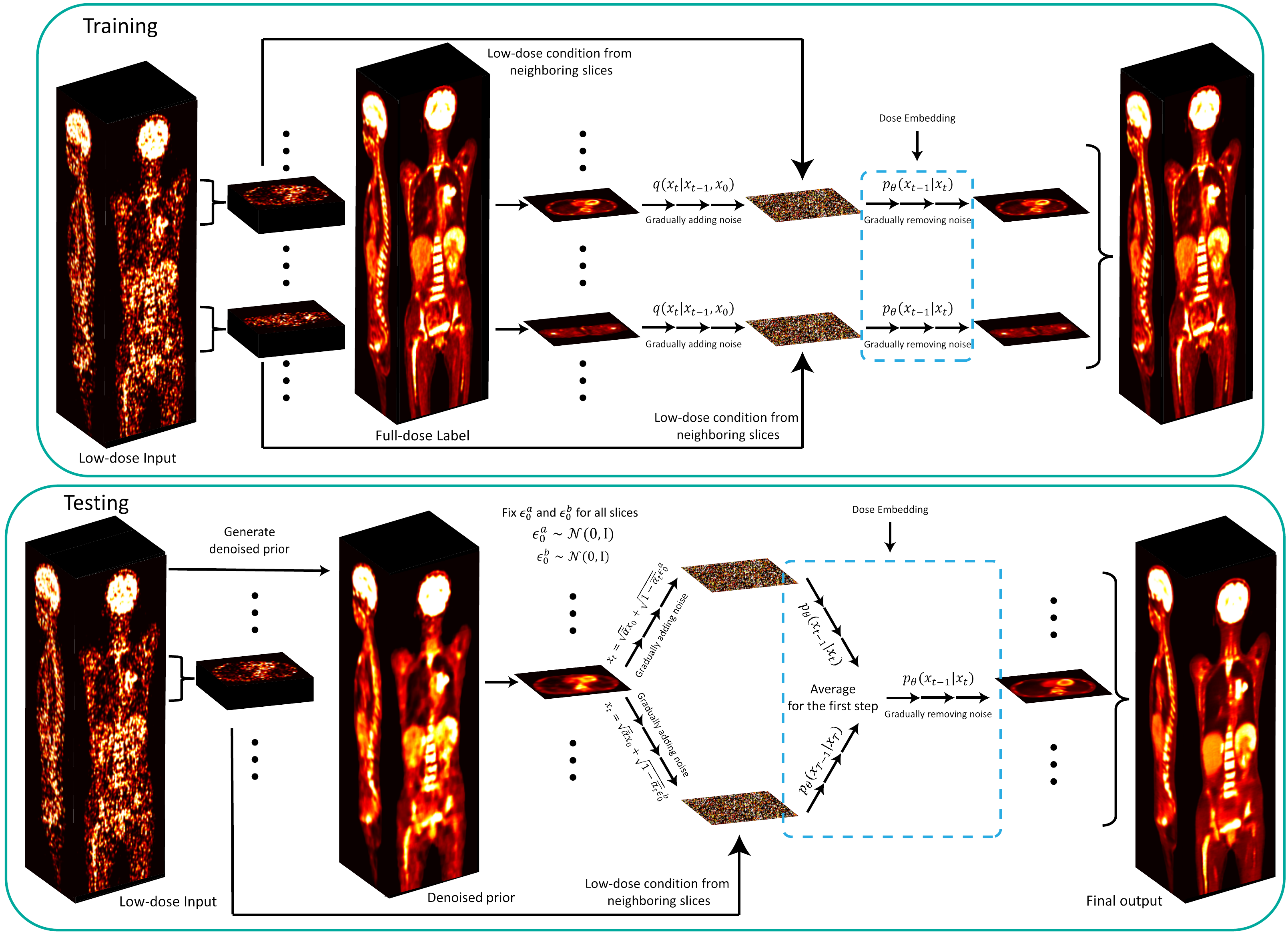}}
\caption{The training (\emph{top}) and the sampling (\emph{bottom}) pipeline of the proposed DDPET-3D method. The DDPET-3D uses multiple neighboring slices as additional inputs to predict the central slice, allowing the network to observe 3D information during training and testing. The DDPET-3D is also conditioned on the injected dose in order to accommodate inputs with varying noise levels. In sampling, we propose to fix the Gaussian latent for each slice in the 3D volume to address the inconsistency along the z-axis. DDPET-3D also uses a pre-trained denoiser prior to ensure accurate quantification in sampling.}
\label{fig_network}
\end{figure*}

The proposed framework is depicted in Fig.~\ref{fig_network}. The proposed network can observe 3D information from adjacent slices during the training process. This is achieved by using the $n$ neighboring slices as input to predict the central slice. Ablation studies showed that the performance converges at $n=31$, which was used in this work. One may use 3D convolutional layers to replace the 2D convolutional layers in the diffusion model for an enlarged receptive field and to allow the network to observe 3D structural information. However, we found that it significantly increases the memory burden and makes the network difficult to optimize. To alleviate the memory burden and allow faster convergence, we embed neighboring slices in the channel dimension. Specifically, when using only one 2D slice as conditional information ($n=1$), the input dimension is $N_b\times W\times W\times2$, where the last dimension is the channel dimension, $N_b$ is the input batch size, and $W$ is the width of the images. One channel is a conditional 2D slice, and the other one is $\xbm_{t-1}$. When $n=31$, the input dimension becomes $N_b\times W\times W\times32$. 

This technique allows the network to observe neighboring slices for 3D image reconstruction with only an incremental increase in memory burden. In addition, ablation studies showed that this technique also noticeably improves the inconsistency problem for 3D imaging. This is consistent with our expectation. With $n=1$, when the network tries to predict the next slice, we observe inconsistent reconstructions because both the conditional information and the starting Gaussian noise $p(\xbm_T)$ change. When $n=31$, the starting Gaussian noise $p(\xbm_T)$ still changes, but all the 31-channel conditional information only shifts a little. The network should produce a more consistent output with only subtle changes in the input.

We hypothesized that another reason for the inconsistency issue is that the starting Gaussian noise of the reverse process is different for different slices. The diffusion sampling strategy starts from a random location in the high-dimensional space (random Gaussian noise), and approximates the data distribution $q(\xbm_0)$ based on the trained neural network, or the score function as described in \cite{song_score-based_2021}, through many iterations. The denoising problem is ill-posed, and we could generate an infinite number of different denoised images given the same low-count input with different starting Gaussian noise. This is beneficial for generative models to produce a wide variety of different images. However, the stochastic nature of the diffusion model would be problematic for medical image reconstruction problems because we expect neighboring slices to be consistent with each other in the 3D image volume. To address this problem, we proposed to fix the starting Gaussian noise when reconstructing all the slices in the entire 3D volume. Specifically, the starting Gaussian noise at the last time step $T$, $\xbm_T$ is fixed for all the slices during the sampling process, only the conditional low-count PET images change for different slices.

This approach produced more consistent reconstructions along the z-axis (as demonstrated in Fig. \ref{fig_results_no_fix_seed_one_channel}). However, we noticed that by fixing $p(\xbm_T)$, there are some noise-dependent artifacts that propagate to all the slices along the z-axis. An example is presented in the first sub-figure in Fig.~\ref{fig_results_no_fix_seed_one_channel} (single $\epsilon$ image). To address this issue, as presented in Fig.~\ref{fig_network}, we proposed to initialize 2 different noise variables ${\bm\epsilon}_0^a$ and ${\bm\epsilon}_0^b$, and we have 2 different $\xbm_{t-1}$ in the reverse process. It is worth noting that having 2 different noise variables does not significantly increase sampling time since we only need to average them at the first reverse step. This technique effectively addresses this issue.

As mentioned in the Introduction, despite the visually appealing results, the diffusion model usually produced images with inaccurate quantification. We noticed that, even though U-net-based methods failed to produce satisfactory results, especially for extremely low-dose settings, they typically maintained overall image quantification much better than diffusion models. To take advantage of that, instead of starting from random Gaussian noise, we first generate a denoised prior using a pre-trained U-Net-based network. Gaussian noise is then added to this prior following Equation \ref{eq3}, and the resulting perturbed image serves as the starting point for the reverse diffusion process. We adapted the Unified Noise-aware Network (UNN) proposed by Xie \textit{et al.} \cite{xie_unified_2023} to generate the denoised priors. 

Intuitively, one can imagine that diffusion models aim to approximate a very narrow distribution centered around real PET images ($q(\xbm_0)$). Starting from a high-dimensional Gaussian space makes it difficult for the model to accurately converge to this narrow distribution. In contrast, U-Net outputs provide a strong initial guess that lies closer to the true PET distribution, thereby significantly improving both visual fidelity and quantitative accuracy. This benefit is clearly demonstrated in Fig. \ref{fig_results_no_unet} and Table \ref{table3}.

For the embedding of the injected dose, the values of the administered dose were first converted to Becquerel (Bq). The functions $\sin$ and $\cos$ functions were then used to encode the injected dose and add to the diffusion time steps (i.e., $t+\sin(\mathrm{dose}) + \cos(\mathrm{dose})$, where $t$ is the diffusion time-step). The time step $t$ was also encoded using $\sin$ and $\cos$ functions. The encoded values were then fed into two linear layers to generate the embedding with a Sigmoid linear unit (SiLU) in between.

For diffusion sampling, DDIM sampling enables faster convergence. But we noticed that using DDIM alone would tend to produce over-smoothed images. To alleviate this issue, adapted from the DDIM paper \cite{song_denoising_2022}, we proposed to have an interpolated sampling between DDIM and DDPM. We inserted DDPM samplings every 5 steps in between DDIM samplings to prevent the images from becoming over-smoothed. Algorithm \ref{alg1} displays the complete training procedure of the proposed method. Algorithm \ref{alg2} displays the complete sampling procedure of the proposed method.

In summary, compared to previous methods, DDPET-3D has the following contributions: \emph{(1)} 3D reconstructions with fine details and addresses the issue of 3D spatial inconsistency in previous diffusion models. This was achieved by using multiple neighboring slices as conditional information, enforcing the same Gaussian latent for all slices in sampling, and using multiple starting Gaussian noise. \emph{(2)} DDPET-3D preserves quantitative accuracy by using a pretrained denoised prior in sampling. \emph{(3)} DDPET-3D achieves dose-aware denoising. It can be generalized to different low-count levels. \emph{(4)} With all the proposed strategies, DDPET-3D converges within 25 diffusion steps, allowing fast reconstructions. It takes roughly 15 minutes to reconstruct the entire 3D volume, while DDPM sampling takes about 6 hours.

\begin{algorithm}[t]
\caption{Training}\label{alg1}
\begin{algorithmic}
\State \textbf{Repeat}
\State $\xbm_0\sim q(\xbm_0)$ \Comment{Sample single-slice full-dose data}
\State $\xbm_{\mathrm{noisy}}\sim q(\xbm_{\mathrm{noisy}})$ \Comment{Sample multi-slice low-dose data}
\State $\mathrm{dose}_{\xbm_\mathrm{noisy}}$ \Comment{Injected dose of $\xbm_\mathrm{noisy}$} 
\State $t\sim \mathrm{Uniform({1,...,T})}$ \Comment{Sample diffusion time-step}
\State ${\bm \epsilon} \sim \mathcal{N}(0,\mathbf{I})$
\State $\nabla_\theta||{\bm \epsilon}-\epsilon_\thetabm(\sqrt{\bar{\alpha}_t}\xbm_0+\sqrt{1-\bar{\alpha}_t}{\bm \epsilon},t,\xbm_\mathrm{noisy}, \mathrm{dose}_{\xbm_\mathrm{noise}})$ \\ \hfill \Comment{Take gradient descent step}
\State \textbf{Until} convergence
\end{algorithmic}
\end{algorithm}

\begin{algorithm}[t]
\caption{Testing}\label{alg2}
\begin{algorithmic}
\State $\xbm_{\mathrm{noisy}}\sim q(\xbm_{\mathrm{noisy}})$ \Comment{Get entire low-dose image volume}
\State $\mathrm{dose}_{\xbm_\mathrm{noisy}}$ \Comment{Injected dose of $\xbm_\mathrm{noisy}$} 
\State $\xbm_{\mathrm{prior}} = \mathrm{UNN}(\xbm_{\mathrm{noisy}})$ \Comment{Generate denoised prior using} \\ \hfill {a pre-trained network (UNN) \cite{xie_unified_2023}}
\State $\xbm_\mathrm{recon}\gets \mathrm{zeros}(\xbm_{\mathrm{noisy}})$ \Comment{Initialize $\xbm_\mathrm{recon}$}
\State ${\bm \epsilon}_0^a\sim \mathcal{N}(0,\mathbf{I})$, ${\bm \epsilon}_0^b\sim \mathcal{N}(0,\mathbf{I})$ \Comment{Obtain 2 noise variables}
\\
\While{$s=1,...,S$} \Comment{Total number of slices}
\State $\xbm_{\mathrm{noisy}}[s]$ \Comment{Conditioned multi-slice low-dose input}
\State $\xbm_{\mathrm{prior}}[s]$ \Comment{Conditioned single-slice denoised prior}
\State $\xbm_{T'}^a = \sqrt{\bar{\alpha}_{T'}}\xbm_{\mathrm{prior}}[s]+\sqrt{1-\bar{\alpha}_{T'}}{\bm \epsilon}_{0}^a$ 
\State $\xbm_{T'}^b = \sqrt{\bar{\alpha}_{T'}}\xbm_{\mathrm{prior}}[s]+\sqrt{1-\bar{\alpha}_{T'}}{\bm \epsilon}_{0}^b$ 
\\
\While{$t=T',...,1$}
\State $z \sim \mathcal{N}(0,\mathbf{I})$
\State $\xbm_{t-1}^a=\mathrm{Sampler}(\xbm_{t}^a,\xbm_{\mathrm{noisy}}[s],\mathrm{dose}_{\xbm_\mathrm{noisy}})$

\If{$t==T'$}
\State $\xbm_{t-1}^b=\mathrm{Sampler}(\xbm_{t}^b,\xbm_{\mathrm{noisy}}[s],\mathrm{dose}_{\xbm_\mathrm{noisy}})$
\State $\xbm_{t-1}^a = (\xbm_{t-1}^a + \xbm_{t-1}^b)/2$
\EndIf
\EndWhile
\State $\xbm_\mathrm{recon}[s] = \xbm_{1}^a$
\\
\EndWhile
\State \textbf{Return }$\xbm_\mathrm{recon}$
\end{algorithmic}
\end{algorithm}

\subsection{PET data acquisitions and reconstructions}
The collection of the patient datasets in each medical center was approved by the Institutional Review Board (IRB) of each institution. All data in this study were de-identified prior to model training, testing, and reader studies.

\textbf{Shanghai Ruijin Hospital Dataset:} 994 subjects with \textsuperscript{18}F-FDG tracer were included in this dataset. All data were acquired using a United Imaging Healthcare uExplorer total-body PET/CT system. Images were reconstructed using the OSEM algorithm with 4 iterations and 20 subsets. A $5 \si{mm}$ FWHM Gaussian filter was applied after reconstructions. The reconstruction matrix size was $673 \times 360\times 360$ with a $2.89\times 1.67 \times 1.67 \si{mm}^3$ voxel size. We randomly selected 210 subjects for training, 10 subjects for validation, and 774 subjects for testing. The average full administered activity (in Megabecquerel) was $217.4\pm53.3\ \mathrm{MBq}$ with scan duration of 5 min and post-injection tracer uptake time of 60 min. 1\%, 2\%, 5\%, 10\%, 25\%, and 50\% low-count levels were available through listmode rebinning. 50\% low-count images were available only for 320 subjects. \textbf{Real Low-dose Dataset: }20 real low-dose scans were also acquired at Shanghai Ruijin Hospital with average administered activity of $27.1\pm5.4\ \mathrm{MBq}$.

\textbf{University Hospital of Bern Dataset: }377 subjects with the \textsuperscript{18}F-FDG tracer were included in this dataset. All data were acquired using a Siemens Biograph Vision Quadra total-body PET/CT system. Images were reconstructed using OSEM with 6 iterations and 5 subsets. A $5 \si{mm}$ FWHM Gaussian filter was applied after reconstructions. The reconstruction matrix size was $644\times 440\times 440$ with a $1.65 \times 1.65 \times 1.65 \si{mm}^3$ voxel size. We randomly selected 99 subjects for training, 10 subjects for validation, and 268 subjects for testing. The average full administered activity was $218.7\pm49.4\ \mathrm{MBq}$ with scan duration of 10 min and post-injection tracer uptake time of 60 min. 1\%, 2\%, 5\%, 10\%, 25\%, and 50\% low-count levels were available through listmode rebinning.

\textbf{Yale-New Haven Hospital Dataset: }195 subjects with the \textsuperscript{18}F-FDG tracer were included in this dataset. All data were acquired using a Siemens Biograph mCT PET/CT system. The whole-body scan protocol with continuous bed motion scanning was used. Images were reconstructed using OSEM with 2 iterations and 21 subsets. A $5 \si{mm}$ FWHM Gaussian filter was applied after reconstructions. The voxel size of the reconstructed image was $2.04\times2.04\times2.03 \si{mm}^3$. We randomly selected 90 subjects for training, 10 subjects for validation, and 95 subjects for testing. The image size was $400 \times 400$ in the transverse plane and varied in the axial direction depending on patient height. The average full administered activity was $256.3\pm16.2 \mathrm{MBq}$ with scan duration of 5 min every bed position and post-injection tracer uptake time of 60 min. 5\%, 10\%, and 20\% low-count levels were available through listmode rebinning.

\textbf{UC Davis Medical Center Dataset: }10 subjects with the \textsuperscript{18}F-FDG tracer were included in this dataset. All data were acquired using a United Imaging Healthcare uExplorer total-body PET/CT system. Images were reconstructed using the OSEM algorithm with 4 iterations and 20 subsets. No post-smoothing was applied. The reconstruction matrix size was $828 \times 256\times 256$ with a $2.34 \times 2.34 \times 2.34 \si{mm}^3$ voxel size. The average full administered activity was $300.6\pm9.0\ \mathrm{MBq}$ with scan duration of 20 min and post-injection tracer uptake time of 120 min. 2.5\%, 6.25\%, 12.5\%, 25\%, and 50\% low-count levels were available through listmode rebinning. Images were interpolated to match the voxel size of Shanghai Ruijin Hospital data when applying the trained network.

All the images were reconstructed using vendor software provided by United Imaging Healthcare and Siemens Healthineers. All the reconstruction parameters are the same as those used in clinical settings in the respective hospitals.
\section{Results}
\subsection{Reader study on three commercial scanners}

\begin{figure*}[!t]
\centerline{\includegraphics[width=0.8\textwidth]{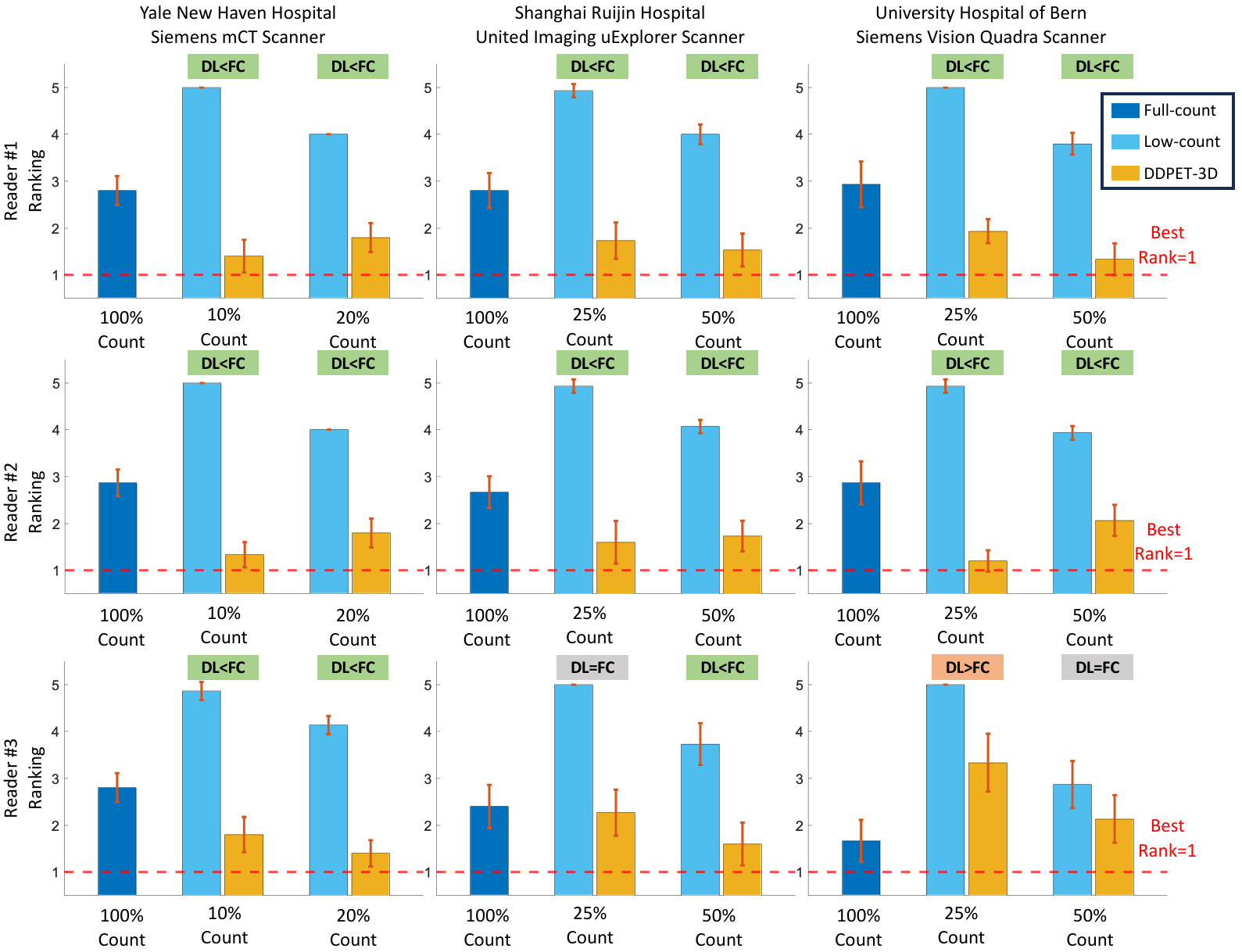}}
\caption{Comparison of rankings by three readers (\#1-3) among images reconstructed with different count-levels and the corresponding denoised images using the proposed DDPET-3D method across three hospitals (Institution \#1-3). Error bars indicate the 95\% confidence interval. Readers were asked to rank the images based on their overall quality. Red dashed lines indicate the best possible rank (i.e., 1). Compared with 100\% full-count images, the text in the box above each plot gives the statistical testing results by a paired t-test at $p<0.05$. Light red, green, and gray boxes indicate that the DDPET-3D results are worse than, better than, or comparable to the full-count images at 5\% significant level, respectively. FC: full-count; DL: the proposed deep learning method (DDPET-3D).}
\label{reader_study_3_sites}
\end{figure*}

\begin{figure*}[!t]
\centerline{\includegraphics[width=0.8\textwidth]{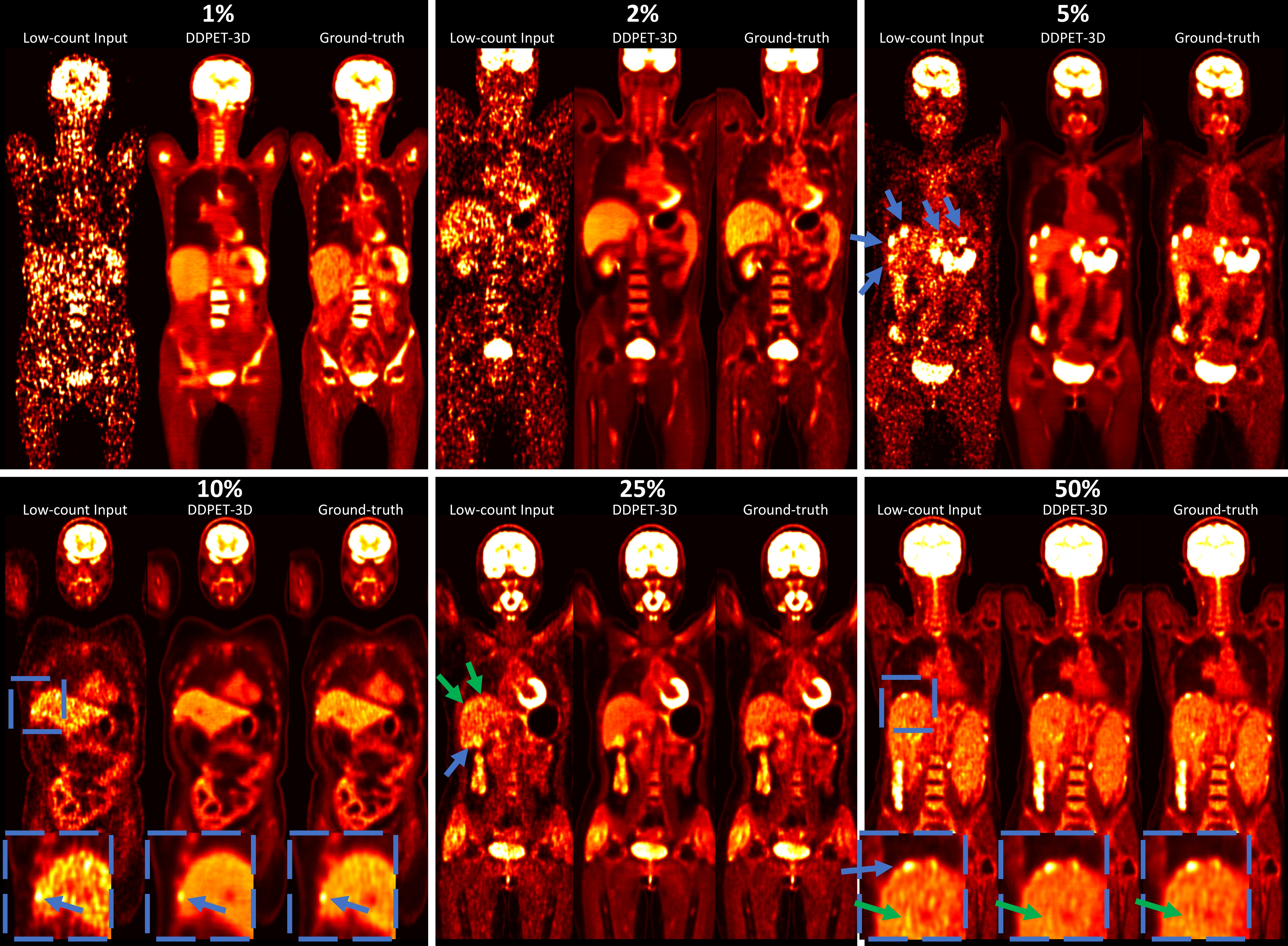}}
\caption{Sample low-count denoised images reconstructed using the proposed DDPET-3D with different low-count levels. 6 different patients at different low-count levels were randomly chosen from the testing dataset. DDPET-3D can produce consistent denoised results for a wide range of different noise-levels. Ground-truth images were reconstructed using 100\% full-count data.  As determined using the 100\% full-count images, blue arrows point to true lesions or hot spots in different patients, and green arrows point to false hot spots in the images. Blue dashed boxes indicate zoom-in ROI.}
\label{UI_patient_all_doses}
\end{figure*}

We performed a reader study to evaluate the network performance. 45 patient studies acquired using three different commercial scanners from three hospitals (15 from each hospital) were randomly selected from the testing dataset and included for the reader study. The three types of commercial scanner are the Siemens mCT scanner, the Siemens Vision Quadra scanner, and the United Imaging uExplorer scanner. Three hospitals are Yale New Haven Hospital in the US, Shanghai Ruijin Hospital in China, and the University Hospital of Bern in Switzerland.

Low-count images were reconstructed by down-sampling PET listmode events. For images acquired on Siemens Vision Quadra and United Imaging uExplorer scanners, six low-count levels were available, including 1\%, 2\%, 5\%, 10\%, 25\%, and 50\%. For images acquired using the Siemens mCT scanner, three low-count levels were available, including 5\%, 10\%, and 20\%. The images were reconstructed using vendor software from United Imaging Healthcare and Siemens Healthineers. Details on image reconstruction are provided later in this paper (see Methods). 

Three board-certified nuclear medicine physicians from the Yale-New Haven Hospital participated independently in the study of the readers. Five 3D images of each patient were provided to the readers simultaneously. The readers were asked to rank the presented images based on the overall image quality. To simulate clinical reality, the images were presented to the readers in DICOM format, and they were able to choose different display views and adjust the display scale if necessary. 

We chose images from the two higher low-count levels for the reader study, since they would be more practical to achieve in clinical settings. For patient studies acquired on the Siemens Vision Quadra and the United Imaging uExplorer scanners, five images are the 25\% and 50\% low-count images, the two corresponding denoised images using the proposed DDPET-3D method, as well as the 100\% full-count image. For the patient studies acquired on the Siemens mCT scanner, five images are the 10\% and 20\% low-count images, the two corresponding denoised images using the proposed DDPET-3D method, as well as the 100\% full-count image. 

The reader study results are summarized in Fig. \ref{reader_study_3_sites}. For data acquired using the Siemens mCT scanner, all three readers preferred the 10\% and 20\% denoised results generated by the DDPET-3D method over the 100\% full-count images with overall superior average rankings. 

For the United Imaging uExplorer data, readers \#1 and \#2 preferred the denoised images over the 100\% full-count images. Reader \#3 gave similar rankings between 100\% full-count images and the 25\% denoised images ($p=0.67$), while rankings of 50\% denoised images is still higher than 100\% full-count images with statistical significance ($p<0.05$).

For the Siemens Vision Quadra data, readers \#1 and \#2 preferred the denoised images over the 100\% full-count images. Reader \#3 gave overall lower rankings for the 25\% denoised images than the 100\% full-count images, while at 50\% count-level, reader \#3 gave DDPET-3D and 100\% full-count images similar rankings.

The three readers agreed that the proposed DDPET-3D can generate denoised results that are better or at least comparable to the 100\% full-count images in terms of overall image quality across the three types of commercial scanners at different hospitals. Sample denoised images acquired at the Shanghai Ruijin Hospital (United Imaging uExplorer scanner) are presented in Fig. \ref{UI_patient_all_doses}. DDPET-3D can generate consistent denoised results across a wide range of low-count levels. DDPET-3D was able to recover fine details in the images (e.g., rib bone in 1\% results; myocardium, spine, and femur in 2\% results). DDPET-3D also effectively suppressed background noise (e.g., liver in patients shown in Fig. \ref{UI_patient_all_doses}), and the lesions became clearer in the denoised images (blue arrows in Fig. \ref{UI_patient_all_doses}). DDPET-3D effectively removed false hot spots (green arrows in the 25\% results) in the images, and the true lesions/hot spots are easier to identify. True lesions/hot spots and false hot spots were determined based on the 100\% full-count images. Please note that we are not claiming perfect reconstruction at extremely low count levels (e.g. 1\% or 2\%). These examples are included to illustrate the practical limit and robustness of DDPET-3D under challenging conditions.

\subsection{Reader study on real low-dose scans}

\begin{figure*}[!t]
\centerline{\includegraphics[width=0.8\textwidth]{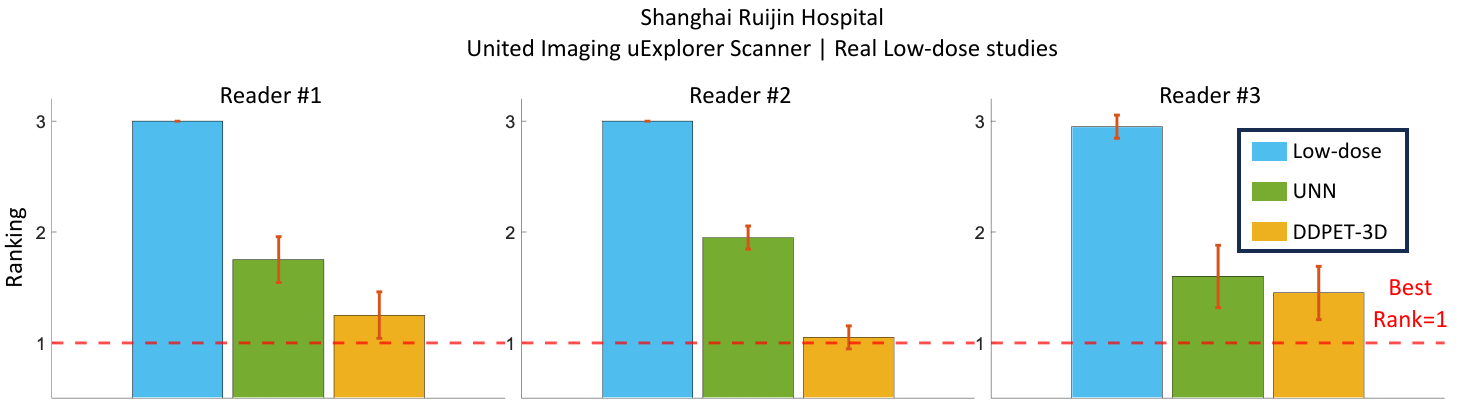}}
\caption{Comparison of rankings of images with low injected dose as well as the corresponding denoised images using the proposed DDPET-3D and the UNN methods across three readers (\#1-3). Error bars indicate the 95\% confidence interval. Readers were asked to rank the images based on their overall quality. Red dashed lines indicate the best possible rank (i.e., 1). Compared with low-dose input, both DDPET-3D and UNN produced images with superior average rankings with statistical significance at $p<0.05$. UNN: Unified Noise-aware Network \cite{xie_unified_2023}.}
\label{reader_study_real_LD}
\end{figure*}

\begin{figure}[!t]
\centerline{\includegraphics[width=\linewidth]{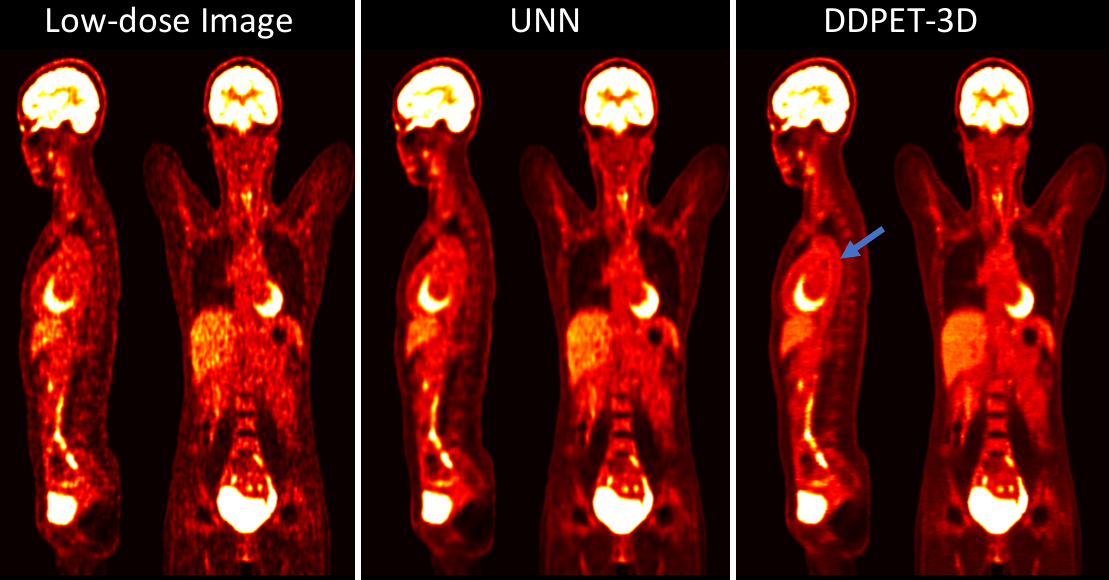}}
\caption{Sample real low-dose denoised images generated using the proposed DDPET-3D and the UNN\cite{xie_unified_2023} methods. Blue arrow points to the aorta wall, which was better reconstructed in DDPET-3D. No full-dose images are available for low-dose scans.}
\label{real_low_dose_UI}
\end{figure}

While rebinning the PET listmode events to generate low-count data to simulate a low-dose setting is a common practice in the literature \cite{xie_unified_2023, zhou_federated_2023, zhou2023fedftn, xu_200x_2017, zhou2020supervised, ouyang_ultra-low-dose_2019, zhou2021mdpet, zhou2023fast, liu2022personalized, gong_pet_2019, onishi_anatomical-guided_2021, liu_pet_2022}, it may affect the random photon rates. To further demonstrate the clinical potential, 20 real low-dose human studies were acquired using a United Imaging uExplorer scanner with an average injected dose of $27.1\pm5.4\text{MBq}$ (about 10\% of full administered dose). Another reader study was performed to assess the image quality. For a more comprehensive comparison, the denoised results generated by the Unified Noise-aware Network (UNN) \cite{xie_unified_2023} were included in the reader study. UNN was chosen because: (1) similar to DDPET-3D, it also achieves noise-aware denoising; (2) it was among the top 10 winning methods in the 2022 Ultra Low-dose PET Imaging Challenge held at the 2022 IEEE Medical Imaging Conference \footnote{https://ultra-low-dose-pet.grand-challenge.org/leaderboard/}.

The reader study was conducted in the similar way as described previously. Three images were presented to the readers simultaneously and they were asked to rank the images based on their overall image quality. Three images are the real low-dose image, denoised images generated by UNN, and denoised images generated by DDPET-3D. As presented in Fig. \ref{reader_study_real_LD}, all three readers agreed that both DDPET-3D and UNN produced superior denoised results compared to low-dose images. DDPET-3D produced images with overall better rankings than the UNN with statistical significance ($p<0.05$) observed in reader \#1 and \#2. Even though DDPET-3D also had a overall better ranking in reader \#3, $p=0.20$ was observed.

Sample real low-dose denoised images are presented in Fig. \ref{real_low_dose_UI}. In this patient, no obvious lesion was identified. Compared to other images, DDPET-3D effectively suppressed excessive noise (e.g., in the liver) and produced images with clearer organ boundaries (e.g., the aorta wall pointed by the blue arrow in Fig. \ref{real_low_dose_UI}).

\subsection{Comparison with other baseline denoising methods}

\begin{figure*}[!t]
\centerline{\includegraphics[width=\textwidth]{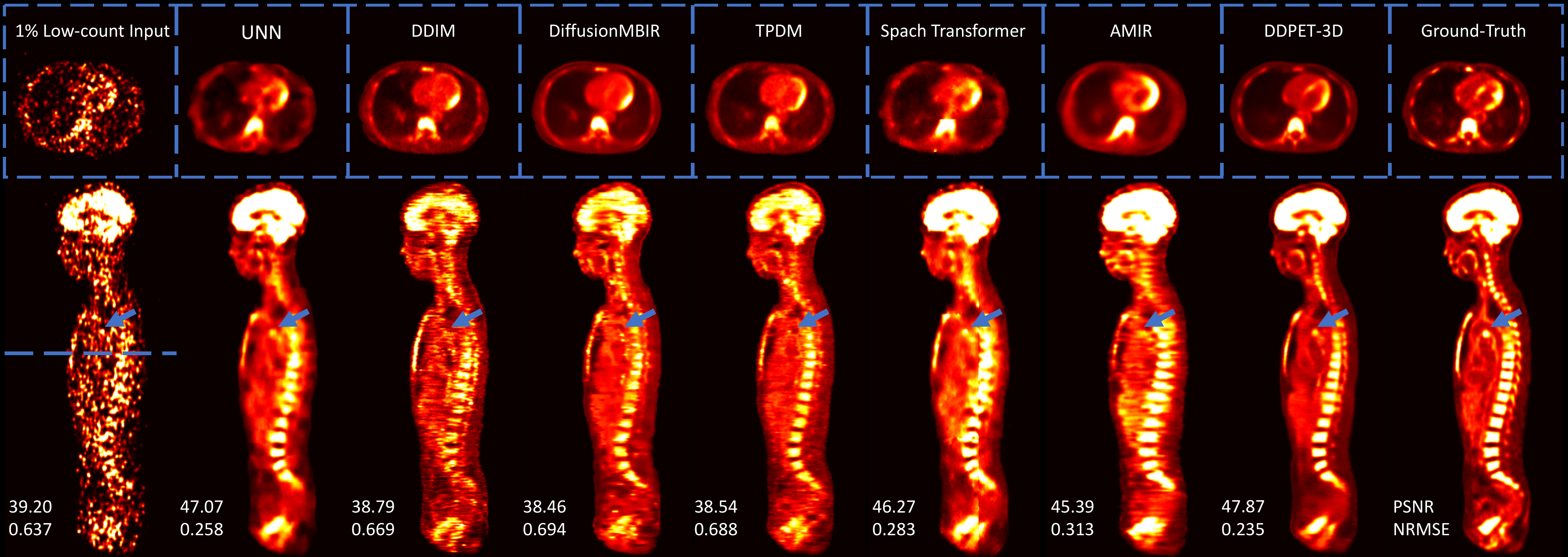}}
\caption{Comparison between DDPET-3D and other baseline denoising methods. Blue arrows point to a hot spot that was better recovered by the proposed DDPET-3D. Dashed blue line indicate the location of the transverse slice. All the images were displayed on the same scale and normalized by the total injected activities.}
\label{UI_results_compare}
\end{figure*}

\begin{table*}[!h]
\centering
\caption{Quantitative comparison of different methods across all test patients from four hospitals. The reported metrics--PSNR, NRMSE, and SSIM--were computed by averaging values over the entire testing dataset. The proposed method consistently produced promising denoised results regardless of input count levels. The best results among different low-count levels are marked in \textcolor{red}{red}. Arrows $\uparrow$ and $\downarrow$ indicate that higher and lower values are better, respectively, for each metric. Asterisks (*) denote statistically significant differences ($p < 0.05$) between the comparison methods and the proposed DDPET-3D.}
\resizebox{\textwidth}{!}{
\begin{tabular}{c|c|c|c|c|c|c}
\hline\hline
\multicolumn{7}{c}{\textbf{Shanghai Ruijin Hospital, United Imaging uExplorer Scanner (674 patients $\times$ 5 $+$ 100 patients $\times6$ = 3,970 studies)}}\\
\hline\hline
     \textbf{PSNR$\uparrow$/NRMSE$\downarrow$/SSIM$\uparrow$} &  1\% Count Input&   2\% Count Input&    5\% Count Input&  10\% Count Input& 25\% Count Input& 50\% Count  Input\\

\hline
Input  & 42.515* / 0.706* / 0.738* & 47.175* / 0.413* / 0.859* & 51.734* / 0.245* / 0.939* & 54.459* / 0.180* / 0.967* & 57.793* / 0.124* / 0.986* & 59.805* / 0.103* / 0.995* \\
\hline
UNN & 50.757* / 0.270* / 0.946* & 52.736* / 0.216* / 0.966* & 54.658* / 0.175* / 0.978* & 55.763* / 0.155* / 0.983* & 56.940* / 0.137* / 0.988* & 57.327* / 0.138* / 0.992* \\
 \hline
  DDIM & 41.608* / 0.760* / 0.868* & 41.644* / 0.757* / 0.886* & 41.652* / 0.757* / 0.898* & 41.659* / 0.756* / 0.903* & 41.674* / 0.755* / 0.909* &  41.106* / 0.768* / 0.907*\\
 \hline
  DiffusionMBIR & 41.827* / 0.742* / 0.903* & 41.795* / 0.744* / 0.912* & 41.759* / 0.747* / 0.917* & 41.745* / 0.749* / 0.919* & 41.729* / 0.750* / 0.922* &  41.171* / 0.754* / 0.925*\\
 \hline
 TPDM & 41.736*  / 0.749* / 0.887* & 41.768* / 0.746* / 0.902* & 41.736* / 0.749* / 0.910* & 41.690* / 0.753* / 0.910* & 41.728* / 0.750* / 0.917* &  41.192* / 0.753* / 0.921*\\
 \hline
Spach Transformer & 48.814* / 0.335* / 0.940* & 52.381* / 0.223* / 0.968* & 55.615 / 0.154 / 0.979* & 57.385 / 0.135* / 0.982* & 59.385* / 0.116* / 0.989* &  61.995 / 0.075 / \textcolor{red}{0.997}\\
 \hline
AMIR & 45.504* / 0.502* / 0.930* & 46.935* / 0.442* / 0.958* & 47.998* / 0.405* / 0.973* & 48.573* / 0.390* / 0.980* & 49.224* / 0.374* / 0.987* &  49.722* / 0.320* / 0.994*\\
 \hline
\textbf{DDPET-3D (proposed)}  & \textcolor{red}{51.673 / 0.244 / 0.957} & \textcolor{red}{53.782 / 0.193 / 0.972} & \textcolor{red}{56.035 / 0.151 / 0.981} & \textcolor{red}{57.495 / 0.130 / 0.986} & \textcolor{red}{59.749 / 0.102 / 0.991} & \textcolor{red}{62.516 / 0.071 / 0.997} \\

\hline
\multicolumn{7}{c}{}\\
\hline\hline
\multicolumn{7}{c}{\textbf{University Hospital of Bern, Siemens Biograph Vision Quadra Scanner (268 patients $\times$ 6 = 1,608 studies)}}\\
\hline\hline
     \textbf{PSNR$\uparrow$/NRMSE$\downarrow$/SSIM$\uparrow$} &  1\% Count Input&   2\% Count Input&    5\% Count Input&  10\% Count Input& 25\% Count Input& 50\% Count  Input\\
\hline
Input  & 47.376* / 0.523* / 0.856* & 50.809* / 0.339* / 0.924* & 54.490* / 0.220* / 0.967* & 56.769* / 0.174* / 0.982* & 59.514* / 0.126* / 0.992 & 61.594* / 0.100* / 0.996* \\
\hline
UNN & 55.154* / 0.208* / 0.977* & 56.823* / 0.170* / 0.985* & 58.470* / 0.141* / 0.989* & 58.986* / 0.137* / 0.991* & 60.527* / 0.114* / 0.994* &  61.877* / 0.098* / 0.995*\\
 \hline
 DDIM & 43.639* / 0.765* / 0.888* & 43.728* / 0.758* / 0.904* & 43.759* / 0.755* / 0.911* & 43.784* / 0.753* / 0.913* & 43.825* / 0.749* / 0.917* &  43.852* / 0.747* / 0.920*\\
 \hline
  DiffusionMBIR & 43.807* / 0.751* / 0.910* & 43.826* / 0.750* / 0.919* & 43.830* / 0.749* / 0.922* & 43.840* / 0.748* / 0.922* & 43.862* / 0.747* / 0.924* &  43.873* / 0.746* / 0.926*\\
 \hline
 TPDM & 43.852* / 0.749* / 0.905* & 43.899* / 0.745* / 0.917* & 43.926* / 0.743* / 0.923* & 43.951* / 0.741* / 0.925* & 43.988* / 0.738* / 0.928* &  44.010* / 0.736* / 0.930*\\
 \hline
Spach Transformer & 53.570* / 0.225* / 0.976* & 57.255 / 0.163* / 0.985 & 58.555 / 0.135 / \textcolor{red}{0.991} & 60.017 / 0.116* / 0.992 & 61.232* / \textcolor{red}{0.093} / 0.993* & 64.156 / 0.076* / \textcolor{red}{0.996}\\
 \hline
AMIR & 49.389* / 0.419* / 0.973* & 49.952* / 0.400* / 0.983* & 50.399* / 0.387* / 0.989* & 50.647* / 0.381* / 0.992 & 50.915* / 0.374* / 0.993* &  51.119* / 0.370* / \textcolor{red}{0.996} \\
 \hline
\textbf{DDPET-3D (proposed)}  & \textcolor{red}{55.909 / 0.191 / 0.980} & \textcolor{red}{57.630 / 0.154 / 0.987} & \textcolor{red}{59.370 / 0.127 / 0.991} & \textcolor{red}{60.631 / 0.113 / 0.993} & \textcolor{red}{62.229 / 0.093 / 0.995} & \textcolor{red}{64.237 / 0.074 / 0.996} \\
\hline

\hline
\multicolumn{7}{c}{}\\
\hline\hline
\multicolumn{7}{c}{\textbf{Yale New Haven Hospital, Siemens mCT Scanner (95 patients $\times$ 3 = 285 studies)}}\\
\hline\hline
     \textbf{PSNR$\uparrow$/NRMSE$\downarrow$/SSIM$\uparrow$} &  \multicolumn{2}{c|}{5\% Count Input} &   \multicolumn{2}{c|}{10\% Count Input }&    \multicolumn{2}{c}{20\% Count Input} \\
\hline
Input  & \multicolumn{2}{c|}{44.929* / 0.474* / 0.831*} & \multicolumn{2}{c|}{48.040* / 0.331* / 0.900*} & \multicolumn{2}{c}{51.272* / 0.227* / 0.947*}  \\
\hline
UNN & \multicolumn{2}{c|}{48.975* / 0.291* / 0.933*} & \multicolumn{2}{c|}{50.514* / 0.243* / 0.954*} & \multicolumn{2}{c}{51.970* / 0.205* / 0.969*}\\
\hline
DDIM & \multicolumn{2}{c|}{40.060* / 0.721* / 0.821*} & \multicolumn{2}{c|}{40.138* / 0.715* / 0.835*} & \multicolumn{2}{c}{40.235* / 0.707* / 0.848*}\\
\hline
DiffusionMBIR & \multicolumn{2}{c|}{40.448* / 0.692* / 0.876*} & \multicolumn{2}{c|}{40.462* / 0.690* / 0.882*} & \multicolumn{2}{c}{40.515* / 0.687* / 0.889*}\\
\hline
TPDM & \multicolumn{2}{c|}{39.784* / 0.698* / 0.863*} & \multicolumn{2}{c|}{39.797* / 0.695* / 0.871*} & \multicolumn{2}{c}{39.945* / 0.690* / 0.880*}\\
\hline
Spach Transformer & \multicolumn{2}{c|}{49.788* / 0.275* / 0.950*} & \multicolumn{2}{c|}{51.003* / 0.207* / 0.963*} & \multicolumn{2}{c}{52.281* / 0.188* / 0.975*}\\
\hline
AMIR & \multicolumn{2}{c|}{45.592* / 0.396* / 0.929*} & \multicolumn{2}{c|}{46.557* / 0.362* / 0.954*} & \multicolumn{2}{c}{47.374* / 0.337* / 0.971*}\\
 \hline
\textbf{DDPET-3D (proposed)}  & \multicolumn{2}{c|}{\textcolor{red}{49.950 / 0.262 / 0.953}} & \multicolumn{2}{c|}{\textcolor{red}{51.383 / 0.199 / 0.966}} & \multicolumn{2}{c}{\textcolor{red}{53.056 / 0.165 / 0.976}}  \\
\hline

\hline
\multicolumn{7}{c}{}\\
\hline\hline
\multicolumn{7}{c}{\textbf{UC Davis Medical Center, United Imaging uExplorer Scanner (10 patients $\times$ 5 = 50 studies)}}\\
\hline\hline
     \multicolumn{2}{c|}{\textbf{PSNR$\uparrow$/NRMSE$\downarrow$/SSIM$\uparrow$}} &  2.5\% Count Input &   6.25\% Count Input &    12.5\% Count Input & 25\% Count Input & 50\% Count Input \\
\hline
\multicolumn{2}{c|}{Input}  & 45.410* / 0.310* / 0.871* & 49.331* / 0.197* / 0.939* & 52.178* / 0.141* / 0.969* & 54.719 / 0.105 / 0.985 & \textcolor{red}{56.860 / 0.082 / 0.994} \\
\hline
\multicolumn{2}{c|}{UNN} & 49.538* / 0.193* / 0.957* & 51.222* / 0.157* / 0.971 & 52.192* / 0.140* / 0.978 & 52.950* / 0.128* / 0.983 & 53.370* / 0.122* / 0.986*\\
\hline
\multicolumn{2}{c|}{DDIM} & 37.436* / 0.754* / 0.840* & 37.542* / 0.746* / 0.864* & 37.578* / 0.743* / 0.874* & 37.608* / 0.740* / 0.882* & 37.633* / 0.738* / 0.889*\\
\hline
\multicolumn{2}{c|}{DiffusionMBIR} & 37.583* / 0.742* / 0.877* & 37.636* / 0.738* / 0.891* & 37.656* / 0.737* / 0.897* & 37.668* / 0.736* / 0.902* & 37.674* / 0.735* / 0.905*\\
\hline
\multicolumn{2}{c|}{TPDM} & 37.388* / 0.758* / 0.846* & 37.551* / 0.745* / 0.871* & 37.635* / 0.738* / 0.887* & 37.621* / 0.739* / 0.888* & 37.494* / 0.750* / 0.876* \\
\hline
\multicolumn{2}{c|}{Spach Transformer} & 50.129 / 0.182 / 0.959* & 50.953* / 0.150* / 0.968* & 52.819* / 0.134* / 0.974* & 54.49 / 0.126* / 0.983 & 55.947 / 0.093* / 0.989\\
\hline
\multicolumn{2}{c|}{AMIR} & 43.900* / 0.379* / 0.946* & 44.800* / 0.350* / 0.963* & 45.475* / 0.333* / 0.971* & 46.130* / 0.318* / 0.982* & 46.798* / 0.305* / 0.989\\
 \hline
\multicolumn{2}{c|}{\textbf{DDPET-3D (proposed)}}  & \textcolor{red}{50.134 / 0.179 / 0.962} & \textcolor{red}{51.964 / 0.145 / 0.973} & \textcolor{red}{53.348 / 0.123 / 0.979} & \textcolor{red}{54.772 / 0.102 / 0.987} & 56.607 / 0.085 / 0.991  \\
\hline

\end{tabular}
}
\label{table2}
\end{table*}

Although most existing deep learning methods are not directly applicable to PET image denoising across varying low-count levels, we believe comparisons with other related methods are still beneficial to show the effectiveness of the proposed method. In this subsection, the proposed method is compared with the following methods:

\begin{enumerate}
\item \textbf{Unified Noise-aware Network (UNN)}\cite{xie_unified_2023}: Similar to the proposed DDPET-3D, UNN was designed for noise-aware PET image denoising. It was among the top 10 winning methods in the 2022 Ultra Low-dose PET Imaging Challenge at the IEEE Medical Imaging Conference, serving as a strong U-Net baseline.
\item \textbf{DDIM} \cite{song_denoising_2022}: A straightforward conditional diffusion model was included for comparison. To maintain a sampling speed similar to that of other diffusion models, DDIM sampling was implemented.
\item \textbf{DiffusionMBIR} \cite{chung_solving_2023}: A diffusion-based model also aims to address the 3D inconsistency issue in diffusion model by applying a TV penalty along the Z-axis.
\item \textbf{TPDM} \cite{lee_improving_2023}: Another diffusion-based method that addresses 3D inconsistency by jointly sampling from two perpendicular diffusion models.
\item \textbf{Spach Transformer} \cite{jang2023spach}: To capture long-range information, the Spach Transformer implements a transformer-based approach to achieve PET image denoising.
\item \textbf{AMIR} \cite{yang2024amir}: AMIR is an all-in-one network capable of handling multiple medical imaging tasks (including PET image denoising) using a single unified model.
\end{enumerate}

As presented in Fig. \ref{UI_results_compare}, standard DDIM sampling produced images with severe inconsistencies between slices in the sagittal view. Both DiffusionMBIR and TPMD improved the results, but the problem still exists. Although AMIR is not a diffusion-based method, it also resulted in inconsistent reconstruction since it is a 2D-based method. The proposed DDPET-3D method produced noticeably more consistent reconstructions. Moreover, even though other diffusion models produced visually realistic PET images, they failed to accurately reconstruct the structure of different organs. For example, all other diffusion models produced images with distorted myocardium as shown in the transverse slice in Fig. \ref{UI_results_compare}. UNN, Spach Transformer, and AMIR also failed to accurately recover the shape of the myocardium. In addition, the proposed DDPET-3D method was also able to recover some fine details in the images, such as the hot spot pointed by the blue arrows in Fig. \ref{UI_results_compare}, which is either over-smoothed (in UNN, Spach Transformer, and AMIR) or barely visible (in other diffusion methods). Although DDPET-3D is not perfect, it consistently outperforms the comparison methods in terms of preserving the anatomical structure and recovering clinically relevant details.

Another major issue of the diffusion model is inaccurate image quantification. Note that all the presented images were already normalized by the total injected activities. As shown in Fig.~\ref{UI_results_compare}, the tracer activities in certain organs are completely wrong in other existing diffusion models. For example, brain activities are noticeably lower in the DDIM, DiffusionMBIR, and TPDM results. Such differences may affect certain diagnostic tasks such as lesion detection \cite{schaefferkoetter2015initial}. Even though UNN produced over-smoothed reconstructions, it does not alter the overall tracer activities in different organs. Using UNN output as denoised prior, the proposed DDPET-3D maintained overall image quantification. Also, using UNN as a denoised prior allows the proposed method to recover some subtle features that are almost invisible in the low-count input (blue arrows in Fig.~\ref{UI_results_compare}).

Image quality was quantitatively evaluated using PSNR (peak signal-to-noise ratio), NRMSE (normalized root mean square error), and SSIM (structure similarity index). Quantitative results are presented in Table \ref{table2}. 

As presented in Table \ref{table2}, the proposed method outperformed other methods in most of the six low-count levels in the four hospitals. Although performance improved gradually for all methods with increasing count levels, only the proposed DDPET-3D method consistently produced optimal results. Both the proposed method and UNN are aware of noise, but UNN performed worse than input at higher count levels (25\% and 50\%). Spach Transformer also showed superior quantitative performance at higher count levels, performing comparably to DDPET-3D in certain settings. Also, due to inaccurate image quantification and slice inconsistency, other comparison diffusion models produced images with even worse quantitative results compared to low-count inputs at higher count levels. Additionally, since the AMIR is a 2D method that reconstructs the 3D volume slice-by-slice, it achieved good SSIM performance but showed poor voxel-wise quantification (PSNR and NRMSE) due to slice inconsistency.

\subsection{Generalizability test in a different hospital}
\begin{figure*}[!t]
\centerline{\includegraphics[width=0.8\textwidth]{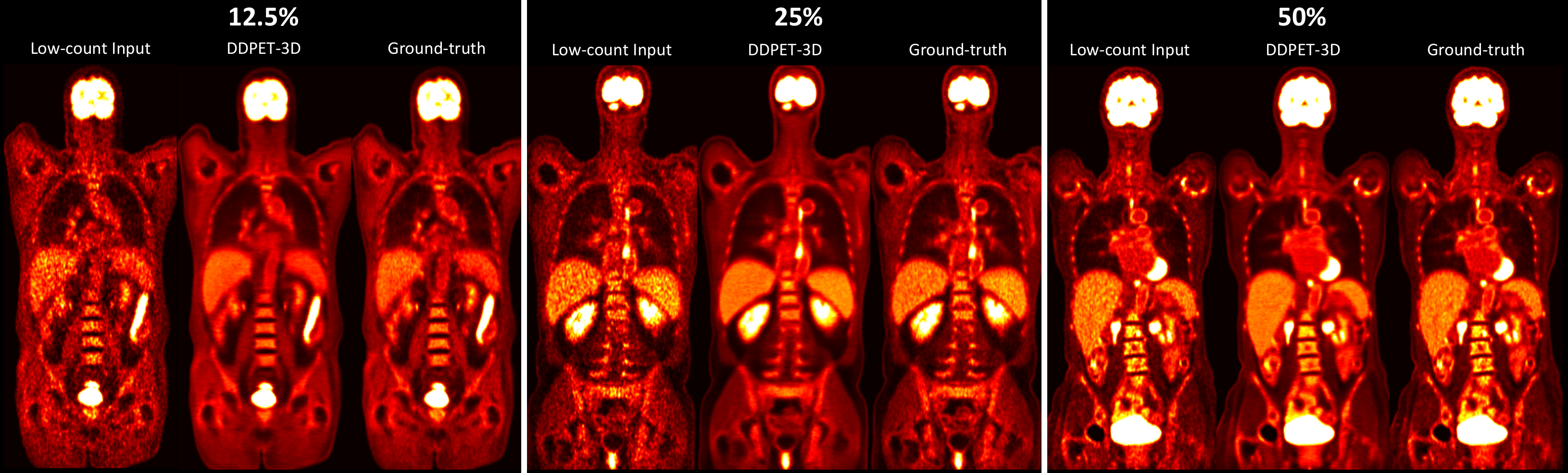}}
\caption{Low-count denoised images reconstructed using the proposed DDPET-3D with different low-count levels for data acquired at the UC Davis Medical Center. The model trained with Shanghai Ruijin data was directly applied for image reconstructions. This illustrates the DDPET-3D may be generalizable to different medical centers without further fine-tuning.}
\label{fig_results_ucd}
\end{figure*}

\begin{figure*}[!t]
\centerline{\includegraphics[width=0.8\textwidth]{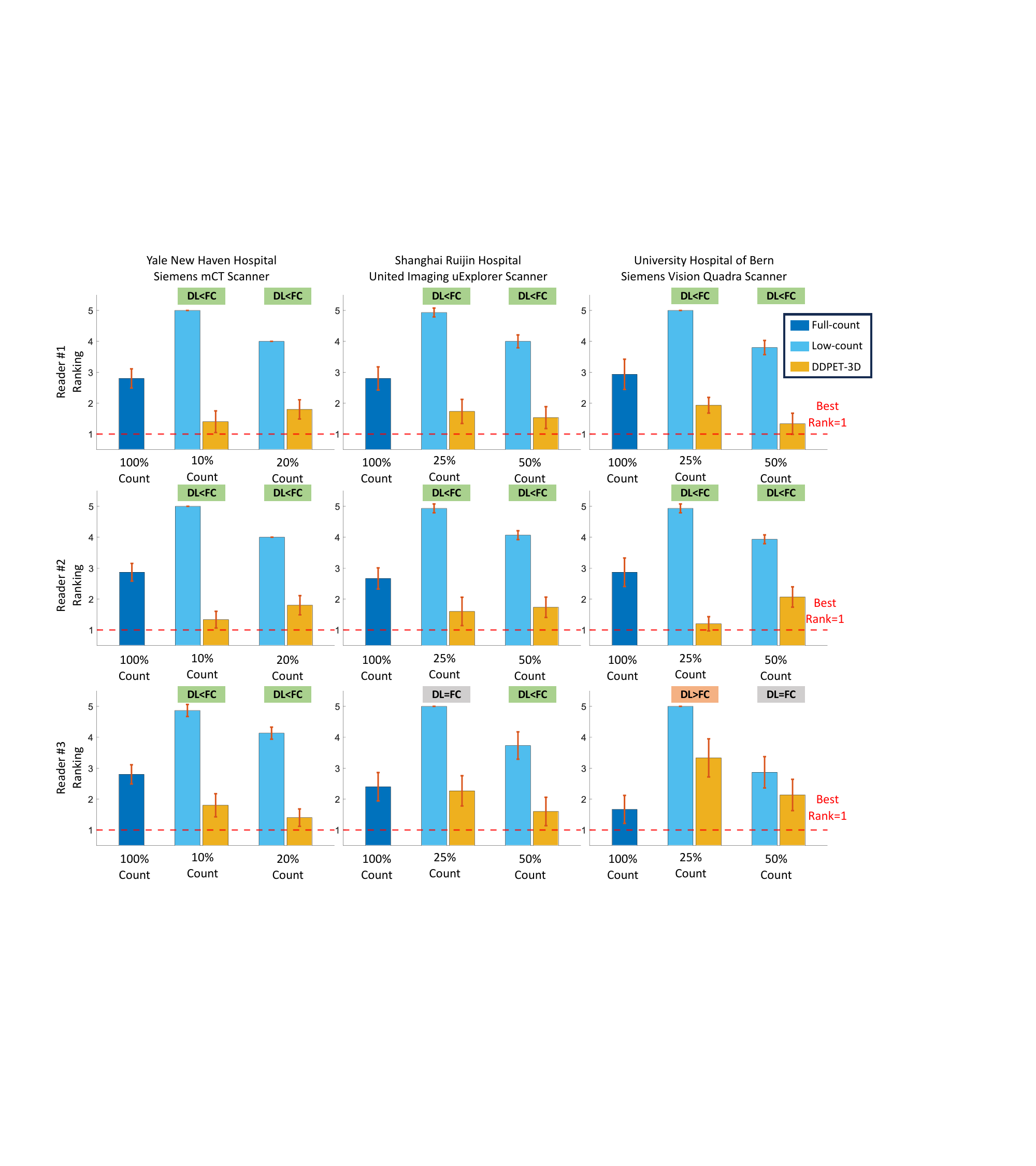}}
\caption{Low-count denoised images reconstructed using the proposed DDPET-3D with different low-count levels for data acquired using a United Imaging uExplorer scanner at the UC Davis Medical Center. The model trained with another United Imaging uExplorer scanner at Shanghai Ruijin data was directly applied for image reconstructions. DDPET-3D can generalize to different medical centers without further fine-tuning. Error bars indicate the 95\% confidence interval. Compared with 100\% full-count images, the text in the box above each plot gives the statistical testing results by a paired t-test at $p<0.05$. Light red, green, and gray boxes indicate that the DDPET-3D results are worse than, better than, or comparable to the full-count images at 5\% significant level, respectively. FC: full-count; DL: the proposed deep learning method (DDPET-3D).}
\label{reader_figure_ucd}
\end{figure*}

\begin{table*}[!t]
\centering
\caption{Quantitative results of the generalizability experiment. The reported metrics--PSNR, NRMSE, and SSIM--were computed by averaging values over the entire testing dataset. Arrows $\uparrow$ and $\downarrow$ indicate that the higher and lower values are better, respectively, for each metric. Asterisks (*) mark statistically significant differences ($p < 0.05$) between the model trained on Ruijin data and the model trained on data from the respective hospital. For ease of comparison, some statistics are copied from Table~\ref{table2}.}
\resizebox{\textwidth}{!}{
\begin{tabular}{c|c|c|c|c|c|c}
\hline\hline
\multicolumn{7}{c}{\textbf{University Hospital of Bern, Siemens Biograph Vision Quadra Scanner (268 patients $\times$ 6 = 1,608 studies)}}\\
\hline\hline
     \textbf{PSNR$\uparrow$/NRMSE$\downarrow$/SSIM$\uparrow$} &  1\% Count Input&   2\% Count Input&  5\% Count Input&  10\% Count Input& 25\% Count Input& 50\% Count  Input\\
\hline
\textbf{DDPET-3D (Trained on Ruijin)}  & 54.590* / 0.264* / 0.965* & 56.267* / 0.222* / 0.975* & 58.215* / 0.150* / 0.985 & 59.470 / 0.143* / 0.989 & 61.259* / 0.115* / 0.993 & 63.650 / 0.087 / 0.996 \\
\hline
\textbf{DDPET-3D (Trained on Bern)}  & 55.909 / 0.191 / 0.980 & 57.630 / 0.154 / 0.987 & 59.370 / 0.127 / 0.991 & 60.631 / 0.113 / 0.993 & 62.229 / 0.093 / 0.995 & 64.237 / 0.074 / 0.996 \\
\hline

\multicolumn{7}{c}{}\\
\hline\hline
\multicolumn{7}{c}{\textbf{Yale New Haven Hospital, Siemens mCT Scanner (95 patients $\times$ 3 = 285 studies)}}\\
\hline\hline
     \textbf{PSNR$\uparrow$/NRMSE$\downarrow$/SSIM$\uparrow$} &  \multicolumn{2}{c|}{5\% Count Input} &   \multicolumn{2}{c|}{10\% Count Input }&    \multicolumn{2}{c}{20\% Count Input} \\
\hline
\textbf{DDPET-3D (Trained on Ruijin)}  & \multicolumn{2}{c|}{49.673* / 0.243* / 0.945*} & \multicolumn{2}{c|}{51.195 / 0.206 / 0.959*} & \multicolumn{2}{c}{53.453* / 0.157* / 0.969*}  \\
\hline
\textbf{DDPET-3D (Trained on Yale)}  & \multicolumn{2}{c|}{49.950 / 0.262 / 0.953} & \multicolumn{2}{c|}{51.383 / 0.199 / 0.966} & \multicolumn{2}{c}{53.056 / 0.165 / 0.976}  \\
\hline
\multicolumn{7}{c}{}\\

\hline\hline
\multicolumn{7}{c}{\textbf{UC Davis Medical Center, United Imaging uExplorer Scanner (10 patients $\times$ 5 = 50 studies)}}\\
\hline\hline
     \multicolumn{2}{c|}{\textbf{PSNR$\uparrow$/NRMSE$\downarrow$/SSIM$\uparrow$}} &  2.5\% Count Input &   6.25\% Count Input &    12.5\% Count Input & 25\% Count Input & 50\% Count Input \\
\hline
\multicolumn{2}{c|}{\textbf{DDPET-3D (Trained on Ruijin)}}  & 50.134 / 0.179 / 0.962 & 51.964 / 0.145 / 0.973 & 53.348 / 0.123 / 0.979 & 54.772 / 0.102 / 0.987 & 56.607 / 0.085 / 0.991  \\
\hline
\multicolumn{2}{c|}{\textbf{DDPET-3D (Trained on Davis)}}  & \diagbox{}{} & \diagbox{}{} & \diagbox{}{} & \diagbox{}{} & \diagbox{}{} \\
\hline
\end{tabular}
}
\label{table_generalize}
\end{table*}

To evaluate the cross-center generalizability of the proposed DDPET-3D, we applied the model trained using the uExplorer data acquired at Shanghai Ruijin Hospital in China to patient studies acquired at the other three hospitals.

\subsubsection{Same Scanner, Different Hospitals}
We first examine the generalizability of the data from UC Davis Medical Center in the USA. Although both hospitals used the same scanner model, there are numerous factors that could lead to generalization challenges.

In addition to different patient characteristics in different countries, two hospitals also implement different reconstruction parameters, reconstruction voxel sizes, and acquisition protocols (see details in Methods). Images acquired at different hospitals also have varied image contrast due to differences in scan time and acquisition start time. Shanghai Ruijin Hospital implemented a total of 5 min of scan duration. At UC Davis hospital, they implemented a total of 20 min of scan duration, similar to a typical multi-bed position whole-body scan. Additionally, UC Davis Medical Center also doubled the conventional tracer uptake time from 60 min post-injection to 120 min post-injection for improved tumor to background ratio in oncologic imaging. Shanghai Ruijin Hospital implemented the conventional 60 min tracer post-injection time.

Ten patient studies from the UC Davis Medical Center were included in this experiment. Five low-count levels were generated through listmodel rebinning, including 2.5\%, 6.25\%, 12.5\%, 25\%, and 50\%. Sample denoised results acquired at the UC Davis Medical Center are presented in Fig. \ref{fig_results_ucd}. The proposed DDPET-3D showed superior generalizability when directly applied to data acquired at a different hospital. The corresponding quantitative measurements are included in Table \ref{table_generalize}.

Another reader study was conducted to evaluate the overall image quality of the 10 UC Davic patients.  Similarly, 25\% and 50\% low-count and denoised images, as well as 100\% full-count images, were included for reader study. As presented in Fig. \ref{reader_figure_ucd}, even though the DDPET-3D denoised images have similar quantitative measurements to low-count images shown in Table \ref{table2} at 25\% and 50\%, all three readers unanimously gave superior overall rankings to 25\% and 50\% denoised results than the 100\% full-count images. Since the quantitative values shown in Table \ref{table2} were calculated using 100\% full-count as reference, they may not correspond to the clinical performance and physicians' preferences. While the DDPET-3D denoised images appear slightly over-smoothed in Fig. \ref{fig_results_ucd} compared to full-count images, probably because the model was not fine-tuned for UC Davis data, the reader study still favored them for their improved noise suppression, demonstrating the superior generalizability of the proposed DDPET-3D. However, further validation against clinical gold standards is needed to fully demonstrate its clinical potential across diverse institutions and imaging protocols.

\subsubsection{Different Scanners, Different Hospitals}
We also applied the trained DDPET-3D model to other hospitals to assess its generalizability. As shown in Table \ref{table_generalize}, despite slight performance degradations, the trained DDPET-3D model demonstrated strong performance across different hospitals. In some cases, the performance differences between DDPET-3D models trained on local hospital data and those trained on data from Shanghai Ruijin hospital were not statistically significant ($p > 0.05$), indicating good generalizability across institutions.

These results suggest that DDPET-3D might be transferable to another medical center without further network fine-tuning, though further investigations are needed.

\subsection{Lesion Detectability Evaluation}
\begin{figure*}[!t]
\centerline{\includegraphics[width=\textwidth]{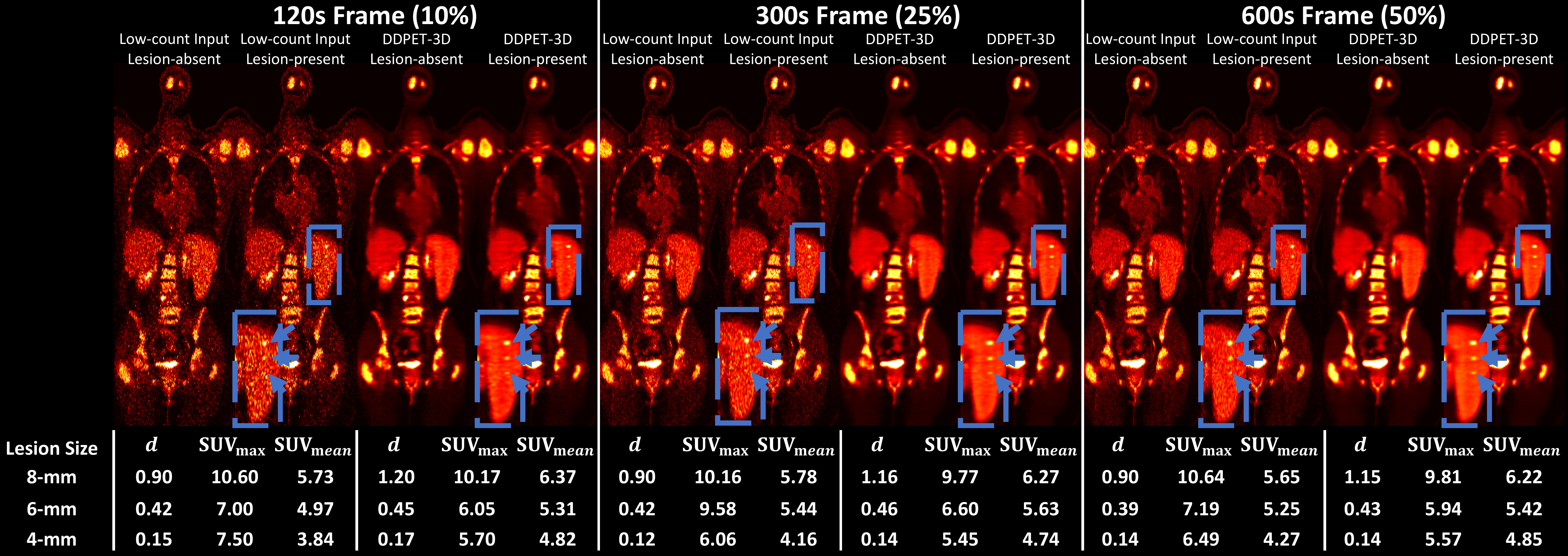}}
\caption{Lesion detectability results on different low-count levels. Arrows point to the three Monte-Carlo simulated lesions at varying sizes. Denoised images at different low-count levels are presented. The lesion detectability index ($d$) and SUV statistics are included in the figure.}
\label{lesion_detect}
\end{figure*}

The previous subsections demonstrated the superior denoising performance of DDPET-3D. Another key challenge in PET image denoising is maintaining lesion detectability while removing excessive noise. In this subsection, we attempt to evaluate lesion quantification before and after denoising.

\subsubsection{Based on Monte-Carlo simulated lesions}Accurately quantifying lesion detectability in vivo is challenging because there is no known ground truth for lesion uptake. To address this issue and evaluate the lesion detectability of the denoised images generated by DDPET-3D, we leveraged a Monte Carlo (MC)-based tool to simulate the listmode data of spherical lesions of varying sizes of 4mm, 6mm, and 8mm in diameter. The lesions were placed in a straight line in the same coronal image slice in the spleen. The simulated listmode data were inserted into the raw data of a lesion-absent 73-year-old female (BMI: 24.8) scanned with the United Imaging uExplorer total-body PET/CT system. An activity concentration ratio of 2.5:1 was simulated with respect to the background. The images were reconstructed using the UC Davis in-house OSEM algorithm with 4 iterations and 13 subsets without point spread function modeling. The reconstructed voxel size was $1.67\text{mm} \times 1.67\text{mm} \times 2.89\text{mm}$.

As detailed in \cite{bayerlein_diana_2024}, spheres in air were simulated using the MC-toolkit SimSET \footnote{\href{https://depts.washington.edu/simset/html/user_guide/user_guide_index.html}{SimSET Homepage}} \cite{poon_validation_2012} with 500 million $^{18}$F decays. SimSET has previously been successfully employed for scatter correction \cite{bayerlein_development_2024} and for image reconstruction algorithms \cite{zhang_quantitative_2017} in total-body PET. Patient attenuation, scatter, detector normalization, and decay during the measurement were taken into account. Based on the background activity concentration in the spleen, a subset of the total number of simulated events was chosen to correctly reflect the desired activity concentration ratio of 2.5:1. The simulated events were then merged with the measured data through random shuffling, creating a realistic listmode dataset.

As shown in Fig. \ref{lesion_detect}, three noisy images were reconstructed at different frame lengths: 120s, 300s and 600s, corresponding to approximately 10\%, 25\%, and 50\% of the full count level, respectively, at the UC Davis Medical Center. The lesion detectability index was quantified using a non-prewhitening matched filter \cite{surti_experimental_2008}. Specifically, the detectability index was calculated as $d=\lambda_1 - \lambda_0$, where $\lambda_i$ is the cross-correlation between the lesion-present ($i=1$) and lesion-absent ($i=0$) images, respectively, and a matched filter $w$. $w$ is defined as the difference between the lesion-present and lesion-absent ground-truth images. As demonstrated by the calculated lesion detectability indices $d$ shown in Fig. \ref{lesion_detect}, the proposed DDPET-3D method effectively improved/maintained lesion detectability for this simulated case. Based on the small volumes of interest (VOI) placed on the simulated lesions, the denoised images generated by DDPET-3D exhibited lower $\mathrm{SUV_{max}}$ and higher $\mathrm{SUV_{mean}}$, indicating effective noise suppression and improved lesion contrast.

\begin{table*}[!h]
\centering
\caption{Absolute percentage difference in lesion-level SUV statistics for different methods across all test patients from four hospitals. Comparisons are made relative to the full-dose reference. The best results among different low-count levels are marked in \textcolor{red}{red}. Asterisks (*) denote statistically significant differences ($p < 0.05$) between the comparison methods and DDPET-3D.}
\resizebox{\textwidth}{!}{
\begin{tabular}{c|c|c|c|c|c|c}
\hline\hline
\multicolumn{7}{c}{\textbf{Shanghai Ruijin Hospital, United Imaging uExplorer Scanner (674 patients $\times$ 5 $+$ 100 patients $\times6$ = 3,970 studies)}}\\
\hline\hline
     \textbf{Lesion SUV Mean / Max Error (\%)} &  1\% Count Input&   2\% Count Input&    5\% Count Input&  10\% Count Input& 25\% Count Input& 50\% Count  Input\\

\hline
UNN & 36.0* / 44.1* & 26.6* / 36.5* & 17.3* / 28.9* & 13.1* / 24.8* & 8.4* / 19.0* & 4.3* / 11.6* \\
 \hline
  DDIM & 62.9* / 58.5* & 58.9* / 58.0* & 54.8* / 55.8* & 53.2* / 55.6* & 50.4* / 53.5* & 47.1* / 54.0* \\
 \hline
  DiffusionMBIR & 64.8* / 64.4* & 60.6* / 63.5* & 56.5* / 62.0* & 54.4* / 60.6* & 51.7* / 59.0* & 48.9* / 58.2* \\
 \hline
 TPDM & 65.1* / 64.7* & 60.6* / 62.3* & 57.0* / 61.3* & 57.4* / 60.6* & 53.3* / 59.2* & 49.5* / 58.0* \\
 \hline
Spach Transformer & 49.0* / 43.4* & 34.5* / 36.1* & 21.3* / 27.1* & 15.1* / 23.1* & 10.3* / 20.1* & 6.2* / 15.4* \\
 \hline
AMIR & 57.9* / 64.9* & 42.4* / 56.4* & 27.6* / 49.9* & 20.3* / 47.2* & 14.1* / 45.2* & 12.7* / 46.4* \\
 \hline
\textbf{DDPET-3D (proposed)}  & \textcolor{red}{33.7 / 39.0} & \textcolor{red}{21.1 / 28.6} & \textcolor{red}{11.4 / 20.4} & \textcolor{red}{8.0 / 17.0} & \textcolor{red}{5.2 / 13.5} & \textcolor{red}{2.9 / 8.7} \\

\hline
\multicolumn{7}{c}{}\\
\hline\hline
\multicolumn{7}{c}{\textbf{University Hospital of Bern, Siemens Biograph Vision Quadra Scanner (268 patients $\times$ 6 = 1,608 studies)}}\\
\hline\hline
     \textbf{Lesion SUV Mean / Max Error (\%)} &  1\% Count Input&   2\% Count Input&    5\% Count Input&  10\% Count Input& 25\% Count Input& 50\% Count  Input\\
\hline
UNN & 22.0* / 27.1 & 13.0* / 16.7 & 6.5 / \textcolor{red}{13.5} & 5.3 / 13.2* & \textcolor{red}{4.0} / 10.8 & 3.5 / \textcolor{red}{8.1} \\
 \hline
 DDIM & 68.7* / 63.0* & 65.8* / 62.4* & 60.2* / 56.3* & 59.4* / 56.1* & 55.3* / 54.3* & 52.9 / 52.5 \\
 \hline
  DiffusionMBIR & 71.9* / 73.4* & 68.5* / 69.3* & 65.6* / 66.3* & 62.6* / 65.3* & 61.5* / 64.8* & 59.0* / 60.7* \\
 \hline
 TPDM & 69.1* / 70.6* & 65.5* / 66.3* & 61.2* / 61.9* & 57.0* / 58.5* & 53.6* / 55.5* & 51.3* / 55.2* \\
 \hline
Spach Transformer & 26.2* / 24.9 & 15.2* / 17.5 & 7.7* / 13.9 & 5.8 / 11.7 & 4.3 / \textcolor{red}{9.8} & 3.3 / \textcolor{red}{8.1} \\
 \hline
AMIR & 37.4* / 51.2* & 27.6* / 46.6* & 19.6* / 43.8* & 16.0* / 42.5* & 13.6* / 41.8* & 12.5* / 41.6* \\
 \hline
\textbf{DDPET-3D (proposed)} & \textcolor{red}{17.6 / 26.2} & \textcolor{red}{11.0 / 16.2} & \textcolor{red}{6.4 / 13.5} & \textcolor{red}{4.9 / 11.6} & 4.2 / 11.6 & \textcolor{red}{3.2 / 8.1} \\
\hline

\hline
\multicolumn{7}{c}{}\\
\hline\hline
\multicolumn{7}{c}{\textbf{Yale New Haven Hospital, Siemens mCT Scanner (95 patients $\times$ 3 = 285 studies)}}\\
\hline\hline
     \textbf{Lesion SUV Mean / Max Error (\%)} &  \multicolumn{2}{c|}{5\% Count Input} &   \multicolumn{2}{c|}{10\% Count Input }&    \multicolumn{2}{c}{20\% Count Input} \\
\hline
UNN & \multicolumn{2}{c|}{14.4* / 20.5*} & \multicolumn{2}{c|}{9.0 / 12.8} & \multicolumn{2}{c}{8.5* / 10.6*}  \\
\hline
DDIM & \multicolumn{2}{c|}{61.1* / 51.8*} & \multicolumn{2}{c|}{56.8* / 50.6*} & \multicolumn{2}{c}{54.9* / 48.0*}  \\
\hline
DiffusionMBIR & \multicolumn{2}{c|}{61.2* / 59.0*} & \multicolumn{2}{c|}{57.9* / 56.3*} & \multicolumn{2}{c}{55.9* / 55.6*}  \\
\hline
TPDM& \multicolumn{2}{c|}{60.6* / 57.3*} & \multicolumn{2}{c|}{59.3* / 58.0*} & \multicolumn{2}{c}{57.3* / 55.2*}  \\
\hline
Spach Transformer & \multicolumn{2}{c|}{20.3* / 21.8*} & \multicolumn{2}{c|}{12.6* / 13.7*} & \multicolumn{2}{c}{7.1* / 10.3*}  \\
\hline
AMIR & \multicolumn{2}{c|}{26.1* / 41.3*} & \multicolumn{2}{c|}{19.3* / 37.8*} & \multicolumn{2}{c}{14.4* / 34.8*}  \\
 \hline
\textbf{DDPET-3D (proposed)}  & \multicolumn{2}{c|}{\textcolor{red}{12.0* / 18.0*}} & \multicolumn{2}{c|}{\textcolor{red}{8.4 / 12.0}} & \multicolumn{2}{c}{\textcolor{red}{5.3 / 8.4}}  \\
\hline

\hline
\multicolumn{7}{c}{}\\
\hline\hline
\multicolumn{7}{c}{\textbf{UC Davis Medical Center, United Imaging uExplorer Scanner (10 patients $\times$ 5 = 50 studies)}}\\
\hline\hline
     \multicolumn{2}{c|}{\textbf{Lesion SUV Mean / Max Error (\%)}} &  2.5\% Count Input &   6.25\% Count Input &    12.5\% Count Input & 25\% Count Input & 50\% Count Input \\
\hline
\multicolumn{2}{c|}{UNN} & 32.1* / 32.2* & 21.8* / 21.7* & 16.7* / 17.1* & 11.1* / 11.1 & 7.4 / 8.1 \\
\hline
\multicolumn{2}{c|}{DDIM} & 64.9* / 54.5* & 59.1* / 56.9* & 56.3* / 51.3* & 50.6* / 38.8* & 47.1* / 38.7* \\
\hline
\multicolumn{2}{c|}{DiffusionMBIR} & 67.1* / 65.1* & 61.1* / 64.1* & 56.8* / 54.5* & 55.0* / 56.2* & 49.5* / 44.1* \\
\hline
\multicolumn{2}{c|}{TPDM} & 68.3* / 70.3* & 63.9* / 65.7* & 58.0* / 58.7* & 58.3* / 54.3* & 59.2* / 51.0* \\
\hline
\multicolumn{2}{c|}{Spach Transformer} & 35.1* / 33.1* & 24.1* / 22.0* & 16.8* / 17.2* & 12.0* / 14.0 & 9.7* / 13.4* \\
\hline
\multicolumn{2}{c|}{AMIR} & 47.7* / 45.9* & 36.6* / 35.4* & 27.0* / 29.4* & 18.7* / 25.7 & 11.1* / 24.2 \\
 \hline
\multicolumn{2}{c|}{\textbf{DDPET-3D (proposed)}}  & \textcolor{red}{29.8 / 23.0} & \textcolor{red}{17.7 / 13.2} & \textcolor{red}{12.2 / 8.8} & \textcolor{red}{5.9 / 9.3} & \textcolor{red}{4.3 / 7.4} \\
\hline

\end{tabular}
}
\label{table_lesion}
\end{table*}

\subsubsection{Based on a lesion segmentation network} To achieve a large-scale evaluation on lesion detectability, we employed an nnU-Net \cite{isensee_nnu-net_2021} model to generate lesion segmentation masks for follow-up lesion SUV analysis. The nnU-Net was first pre-trained using the autoPET dataset \cite{gatidis_results_2024} and then fine-tuned using the lesion segmentation labels annotated by our board-certified nuclear medicine physicians. Due to limited annotation resources, fine-tuning was performed only on data acquired from Yale New Haven Hospital. Despite being fine-tuned solely on Yale data, the nnU-Net generalized well to patient studies from other hospitals in terms of lesion segmentation. The fine-tuned model was subsequently applied to patient studies from all hospitals to automatically segment potential lesions and generate lesion volumes of interest (VOIs). We then compared lesion-level SUV statistics across different methods against the corresponding full-dose reference data. This large-scale evaluation complements the simulation study and further supports the robustness and clinical applicability of our approach. 

The corresponding results on lesion SUV quantification are summarized in Table \ref{table_lesion}. The results demonstrate that the proposed DDPET-3D method achieved consistent and accurate lesion quantification across multiple hospitals, outperforming existing approaches at different low-count levels.

\revend

\subsection{Ablation studies}
\begin{table*}[!h]
\centering
\caption{Quantitative assessment for different ablation studies. The reported metrics--PSNR, NRMSE, and SSIM--were computed by averaging values over the entire testing dataset. The best results among different low-count levels are marked in \textcolor{red}{red}. The second best results are marked in \textcolor{blue}{blue}. Black voxels were removed for the calculations. $\uparrow$ and $\downarrow$ indicate that the higher and lower values are better, respectively, for each metric. Asterisks (*) denote statistically significant differences ($p < 0.05$) between the comparison methods and the proposed DDPET-3D.}
\resizebox{\textwidth}{!}{
\begin{tabular}{c|c|c|c|c|c|c}
\hline\hline
\multicolumn{7}{c}{\textbf{United Imaging uExplorer Scanner (20 patients $\times$ 6 = 120 studies)}}\\
\hline\hline
     \textbf{PSNR$\uparrow$/NRMSE$\downarrow$/SSIM$\uparrow$} &  1\% Count Input&   2\% Count Input&    5\% Count Input&  10\% Count Input& 25\% Count Input& 50\% Count  Input\\

\hline
DDPET-3D (9 slices) & 52.563* / 0.244* / \textcolor{blue}{0.964} & 54.666* / 0.193 / \textcolor{red}{0.977} & 56.194 / \textcolor{red}{0.148} / 0.984 & 57.688 / \textcolor{red}{0.126} / 0.988 & 60.011 / \textcolor{red}{0.097} / \textcolor{red}{0.993} & 62.267* / \textcolor{blue}{0.076} / 0.995 \\
 \hline
DDPET-3D (21 slices)  & 52.502* / 0.246* / 0.960* & 54.509* / 0.198* / 0.974* & 56.574* / 0.160* / 0.981* & 57.949* / 0.139* / 0.985* & 60.137* / 0.111* / 0.990* &  62.338* / 0.088* / 0.994* \\
\hline
DDPET-3D (41 slices)  & 51.866 / 0.238 / \textcolor{blue}{0.964} & 53.926 / \textcolor{red}{0.189} / \textcolor{red}{0.977} & 56.147 / \textcolor{blue}{0.149} / \textcolor{red}{0.985} & 57.618 / \textcolor{blue}{0.127} / \textcolor{red}{0.989} & 59.948 / \textcolor{blue}{0.099} / \textcolor{red}{0.993} &  62.438* / \textcolor{red}{0.075} / 0.995 \\
\hline
DDPET-3D (no prior)  & 41.497* / 0.847* / 0.868* & 41.638* / 0.833* / 0.898* & 41.704* / 0.827* / 0.913* & 41.715* / 0.826* / 0.920* & 41.747* / 0.823* / 0.927* & 41.767* / 0.821* / 0.932* \\
\hline
DDPET-3D (no fix $\epsilon$)  & \textcolor{red}{52.991*} / \textcolor{red}{0.233*} / 0.961* & \textcolor{red}{55.049*} / \textcolor{blue}{0.187*} / 0.973* & \textcolor{red}{57.225*} / 0.150* / 0.980* & \textcolor{red}{58.695*} / 0.129* / 0.983* & \textcolor{red}{61.026*} / 0.102 / 0.988* &  \textcolor{red}{63.916*} / 0.077 / 0.991* \\
\hline
DDPET-3D (no dose)  & 52.817* / \textcolor{blue}{0.237*} / 0.960* & 54.822* / 0.192* / 0.972* & 56.903* / 0.154* / 0.979* & 58.259* / 0.134* / 0.983* & 60.360* / 0.108* / 0.987* &  62.871* / 0.084* / 0.990* \\
\hline
DDPET-3D (single $\epsilon$)  & 52.289* / 0.251* / 0.959* & 54.261* / 0.203* / 0.974* & 56.324* / 0.164* / 0.983* & 57.701* / 0.142* / 0.987* & 59.917* / 0.113* / 0.992* & 62.548* / 0.087* / 0.995* \\
\hline
\textbf{DDPET-3D (proposed)}  & \textcolor{blue}{52.899} / 0.267 / \textcolor{red}{0.965} & \textcolor{blue}{54.937} / 0.215 / \textcolor{red}{0.977} & \textcolor{blue}{57.119} / 0.171 / \textcolor{red}{0.985} & \textcolor{blue}{58.551} / 0.148 / \textcolor{red}{0.989} & \textcolor{blue}{60.916} / 0.103 / \textcolor{red}{0.993} & \textcolor{blue}{63.804} / 0.077 / \textcolor{red}{0.996} \\
\hline
\end{tabular}
}
\label{table3}
\end{table*}

\begin{table*}[!h]
\centering
\caption{Quantitative assessment for different methods. The reported metrics--PSNR, NRMSE, and SSIM--were computed by averaging values over the entire testing dataset. The proposed method consistently produced promising denoised results regardless of input count levels. The best results among different low-count levels are marked in \textcolor{red}{red}. Black voxels were removed for calculations. $\uparrow$ and $\downarrow$ indicate that higher and lower values are better, respectively, for each metric. Asterisks (*) denote statistically significant differences ($p < 0.05$) between the comparison methods and the proposed DDPET-3D.}
\resizebox{\textwidth}{!}{
\begin{tabular}{c|c|c|c|c|c|c}
\hline\hline
\multicolumn{7}{c}{\textbf{United Imaging uExplorer Scanner (20 patients $\times$ 6 = 120 studies)}}\\
\hline\hline
     \textbf{PSNR$\uparrow$/NRMSE$\downarrow$/SSIM$\uparrow$} &  1\% Count Input&   2\% Count Input&    5\% Count Input&  10\% Count Input& 25\% Count Input& 50\% Count  Input\\

\hline
Input  &44.667* / 0.600* / 0.788* & 49.260* / 0.356* / 0.895* & 53.575* / 0.219* / 0.957* & 56.085* / 0.167* / 0.977*
 & 59.346* / 0.115 / 0.991* & 62.253* / 0.082 / 0.996* \\
\hline
UNN & 52.034* / 0.258* / 0.953* & 54.044* / 0.207* / 0.973* & 55.838* / 0.171* / 0.982* & 56.963* / 0.152* / 0.987*
 & 58.570* / 0.128* / 0.991* &  59.885* / 0.111* / 0.994* \\
 \hline
DDIM50  & 42.550* / 0.751* / 0.904* & 42.718* / 0.737* / 0.926* & 42.846* / 0.726* / 0.938* & 42.942* / 0.718* / 0.944* & 43.065* / 0.708* / 0.952* & 43.145* / 0.702* / 0.957*  \\
\hline
DDIM50+2.5D  & 43.636* / 0.665* / 0.910* & 43.851* / 0.648* / 0.924* & 43.856* / 0.648* / 0.930* & 43.829* / 0.650* / 0.932* &  43.805* / 0.652* / 0.936* & 43.820* / 0.650* / 0.938*  \\
\hline
DiffusionMBIR  & 42.590* / 0.748* / 0.913* & 42.747* / 0.734* / 0.934* & 42.868* / 0.724* / 0.944* & 42.960* / 0.717* / 0.950* &  43.079* / 0.707* / 0.957* & 43.156* / 0.701* / 0.961*  \\
\hline
DiffusionMBIR+2.5D  & 42.615* / 0.745* / 0.913* & 42.834* / 0.727* / 0.935* & 42.892* / 0.722* / 0.945* & 42.911* / 0.720* / 0.948* & 42.931* / 0.719* / 0.952* & 42.950* / 0.717* / 0.955*  \\
\hline
TPDM  & 42.691* / 0.739* / 0.906* & 42.810* / 0.729* / 0.929* & 42.939* / 0.718* / 0.940* & 43.051* / 0.709* / 0.946* & 43.202* / 0.697* / 0.954* & 43.291* / 0.690* / 0.959* \\
\hline
TPDM+2.5D  & 42.334* / 0.769* / 0.887* & 42.458* / 0.759* / 0.912* & 42.483* / 0.757* / 0.923* & 42.499* / 0.755* / 0.927* & 42.528* / 0.753* / 0.932* & 42.536* / 0.752* / 0.935*  \\
\hline
\hline
\textbf{DDPET-3D (proposed)}  & \textcolor{red}{52.899 / 0.267 / 0.965} & \textcolor{red}{54.937 / 0.215 / 0.977} & \textcolor{red}{57.119 / 0.171 / 0.985} & \textcolor{red}{58.551 / 0.148 / 0.989} & \textcolor{red}{60.916 / 0.103 / 0.993} & \textcolor{red}{63.804 / 0.077 / 0.996} \\
\hline
\end{tabular}
}
\label{table4}
\end{table*}

\begin{figure}[!t]
\centerline{\includegraphics[width=\linewidth]{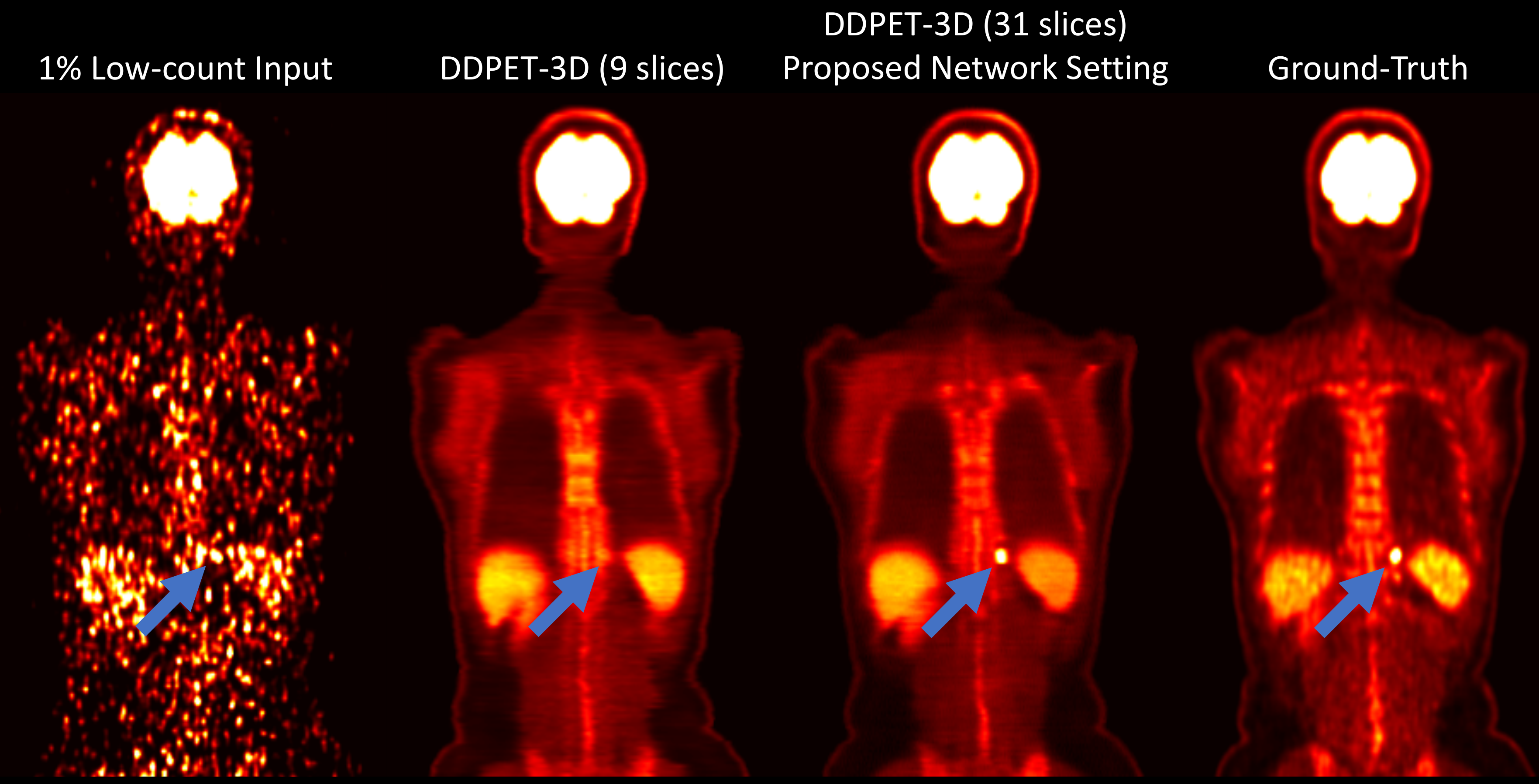}}
\caption{Images reconstructed using DDPET-3D with different numbers of conditional slices. Note that all the images were normalized by the total injected activities. Blue arrows point to a possible lesion that is better recovered by DDPET-3D with more conditional slices.}
\label{fig_results_num_slice}
\end{figure}

\begin{figure*}[!t]
\centerline{\includegraphics[width=0.8\linewidth]{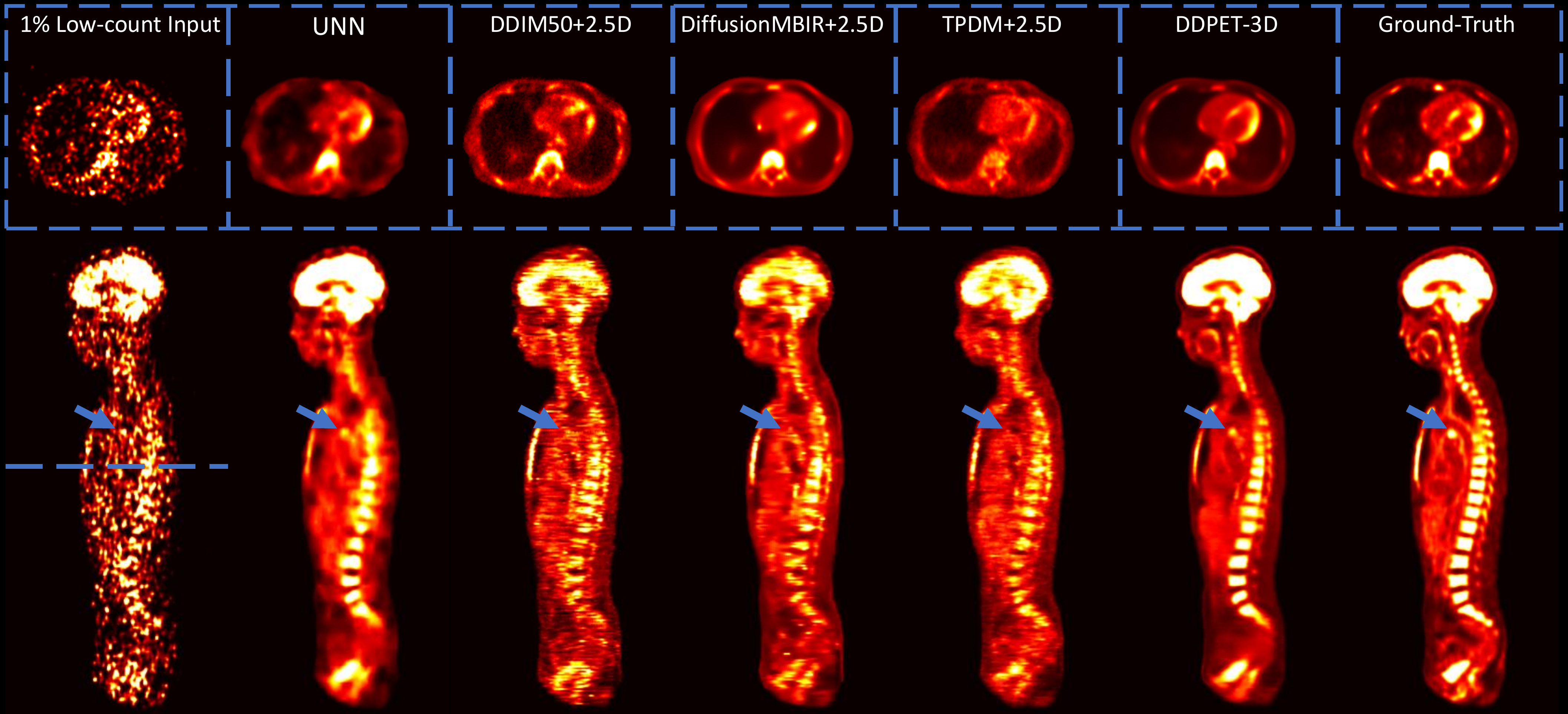}}
\caption{Images reconstructed using different methods with and without the proposed 2.5D conditional strategy. Note that all the images were normalized by the total injected activities. Blue arrows point to a possible lesion that is better recovered by DDPET-3D with more conditional slices.}
\label{fig_results_2_5D_tv_tpmd}
\end{figure*}

\begin{figure}[!t]
\centerline{\includegraphics[width=\linewidth]{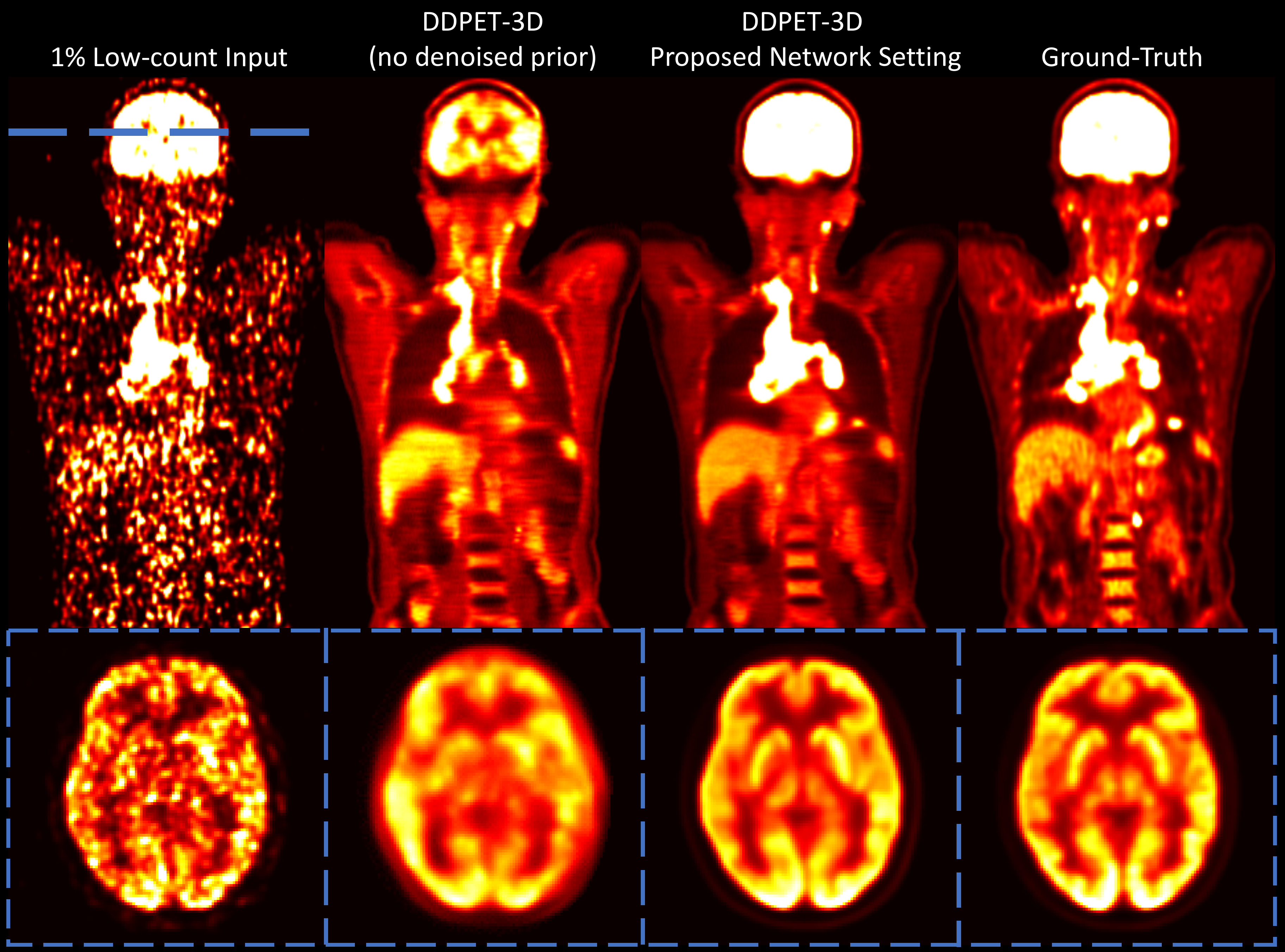}}
\caption{Visual comparison of images reconstructed using the proposed DDPET-3D methods with and without denoised priors. Note that all the images were normalized by the total injected activities. Note the tracer activity differences in different organs. For the brain transverse slice, due to quantification bias, image intensity in the "DDPET-3D (no prior)" method was manually tuned for visual comparison. Many image details were lost with no prior. Dashed blue line indicates the location of the transverse slice.}
\label{fig_results_no_unet}
\end{figure}

\begin{figure*}[!t]
\centerline{\includegraphics[width=0.9\linewidth]{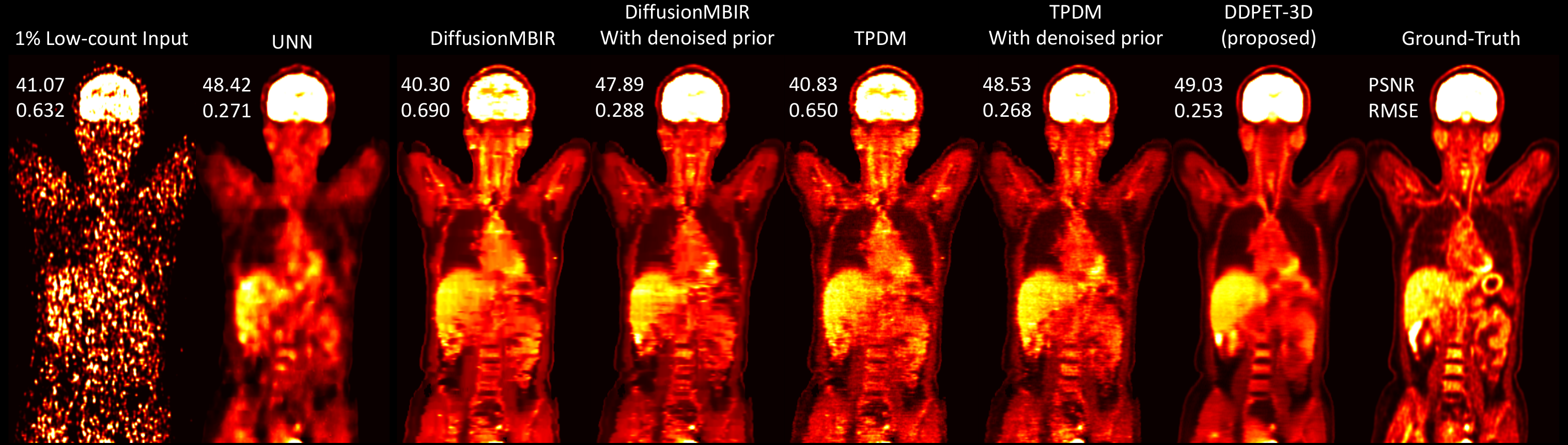}}
\caption{Results produced by competing methods with denoised prior. The proposed DDPET-3D still outperformed other competing methods. Note that the denoised prior is one of the main contributions proposed in this work.}
\label{rebuttal_fig_tv_orth_w_unn}
\end{figure*}

\begin{figure}[!t]
\centerline{\includegraphics[width=\linewidth]{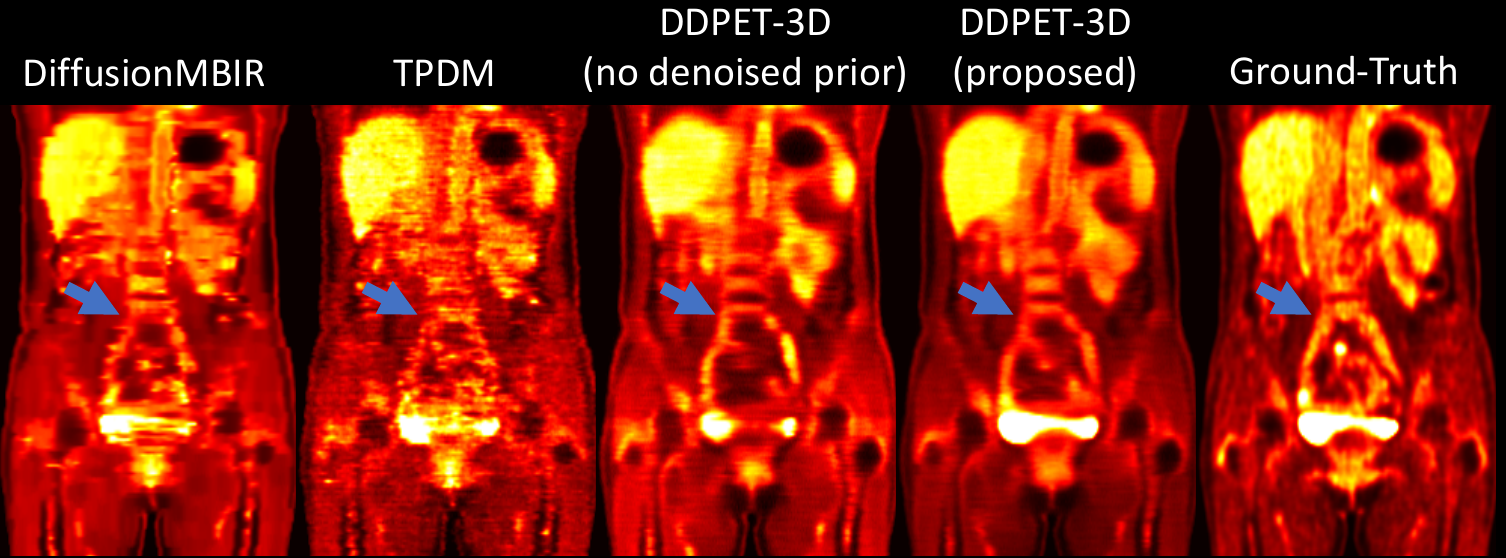}}
\caption{Without the denoised prior, proposed DDPET-3D still produced more consistent reconstructions compared with other methods. Blue arrows point to regions with noticeable inconsistencies between slices in competing methods. Without the denoised prior, DDPET-3D still produced more consistent reconstructions compared with other methods.}
\label{rebuttal_fig_no_unet}
\end{figure}

\begin{figure}[!t]
\centerline{\includegraphics[width=\linewidth]{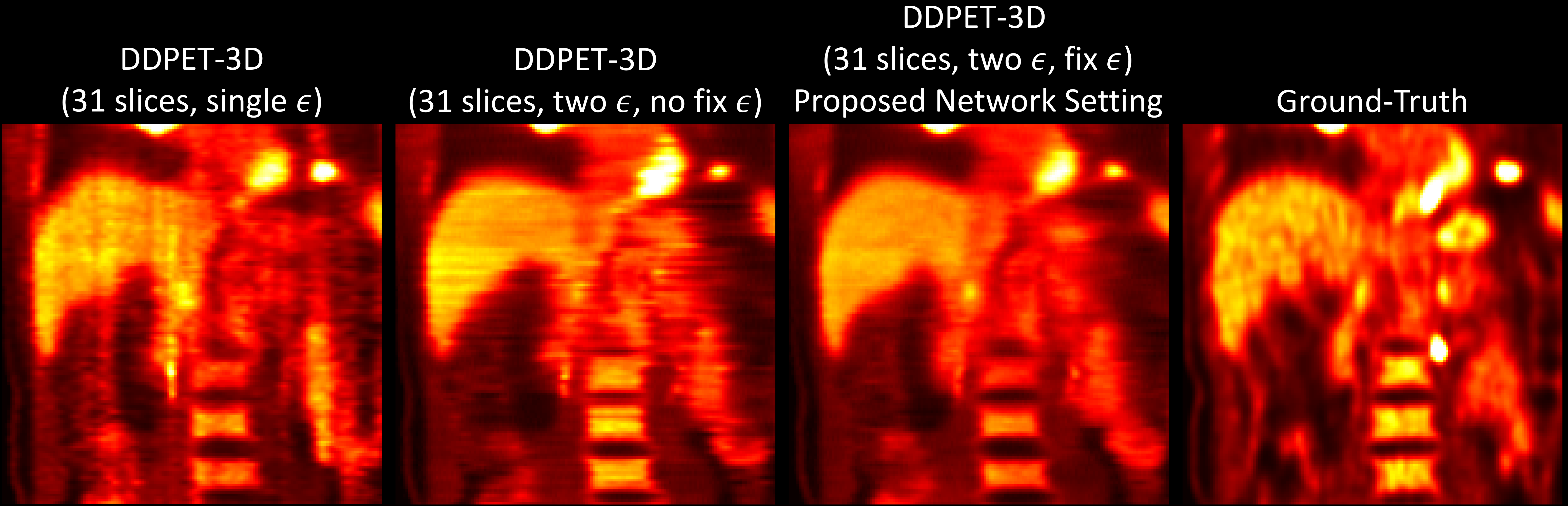}}
\caption{Images reconstructed with different settings of noise variables. Note that all the images were normalized by the total injected activities. Note the undesired artifacts along the z-direction by using only one noise variable $\epsilon$. Also note the inconsistencies of difference slices without fixing $\epsilon$. Despite similar quantitative values, proposed network settings produced images with more consistent 3D reconstructions and less artifacts.}
\label{fig_results_no_fix_seed_one_channel}
\end{figure}

We performed several ablated experiments to demonstrate the effectiveness of different proposed components in the proposed DDPET-3D.

\textbf{Impact of Number of Conditional Slices:} We evaluated DDPET-3D with different numbers of neighboring conditional slices. Three variants of the DDPET-3D networks were trained using 9, 21, and 41 neighboring slices as conditional information. These networks are denoted as "DDPET-3D ($n$ slice)", where $n$ is the number of conditioned neighboring slices. The results showed that $n=31$ performed the best in most quantitative metrics. Therefore, $n=31$ was used in this paper. Quantitative results are presented in Table \ref{table3}. To facilitate the comparison, we only conducted the ablation studies on a subset of the patient studies acquired at the Shanghai Ruijin Hospital. As presented in Fig.~\ref{fig_results_num_slice}, using more conditioned slices helped the network to recover some subtle details in the images (blue arrows in Fig.~\ref{fig_results_num_slice}). 

For comparison, we also extended the comparison diffusion models (DiffusionMBIR, TPDM, and DDIM), and re-trained them using neighboring slices as conditional information (denoted as DiffusionMBIR+2.5D, TPDM+2.5D, and DDIM+2.5D). Sample reconstruction results are presented in Fig. \ref{fig_results_2_5D_tv_tpmd}. Simply adding neighboring slices as conditional information does not necessarily lead to better performance in other baseline models. The corresponding quantitative evaluations are presented in Table \ref{table4}.

\textbf{Impact of Denoised Prior:} We tested the DDPET-3D without the denoised prior during the sampling process to demonstrate the improvement in image quantification using the proposed denoised prior. This network is denoted as "DDPET-3D (no prior)" in Table \ref{table3}. As presented in Fig.~\ref{fig_results_no_unet}, we can see that diffusion models produced images with inaccurate tracer activities in different organs without the proposed denoised prior (especially in the brain and liver). Note that the images were already normalized by the total injected activities of the entire 3D volume. Also, without the denoised prior, many details in the images were not able to be recovered, as presented in the brain slice in Fig.~\ref{fig_results_no_unet}. In addition, we implemented the denoised prior in other competing methods for comparison, so that all the methods have the same starting point in the diffusion process. As shown in Fig. \ref{rebuttal_fig_tv_orth_w_unn}, DDPET-3D still outperformed other competing methods, demonstrating the effectiveness of other proposed techniques in DDPET-3D. With the denoised prior, DiffusionMBIR and TPDM produced images with better PSNR and RMSE measurements than the networks without it, but inconsistencies between slices still exist. Lastly, as shown in Fig. \ref{rebuttal_fig_no_unet}, without the denoised prior, DDPET-3D still produced more consistent reconstructions compared to other methods. Note that the denoised prior is one of the key contributions proposed in this work.

\textbf{Impact of Fixing Noise Variables:} We analyzed the DDPET-3D without and with fixing the 2 noise variables to demonstrate its effectiveness in producing consistent 3D reconstructions. Specifically, $\epsilon_0^a$ and $\epsilon_0^b$ are sampled from $\Ncal(0,\textbf{I})$ for each slice in the 3D image volume, instead of fixing them for all slices. This network is denoted as "DDPET-3D (no fix $\epsilon$)" in Table \ref{table3}. Although the differences in quantitative evaluations without and with fixing noise variables are small, and not fixing noise even led to better quantitative measurements in certain cases, we noticed significant improvements in visual quality, as presented in Fig.~\ref{fig_results_no_fix_seed_one_channel}. Not fixing noise variables produced images with inconsistent slices and unclear organ boundaries, which are unfavorable in clinical settings.

\textbf{Impact of Dose Embedding:} We tested the DDPET-3D without dose embedding, which is denoted as "DDPET-3D (no dose)" in Table \ref{table3}. The experimental results showed that DDPET-3D with dose embedding produced images with better quantitative results. 
\section{Discussion and Conclusion}
We introduced DDPET-3D, a dose-aware diffusion model for 3D low-dose PET imaging. Evaluated on a large-scale, multi-center, and cross-scanner dataset, low-dose/low-count denoised images generated by DDPET-3D performed favorably or comparably relative to the images reconstructed with 100\% full-count. Tested on images acquired using three state-of-the-art PET/CT systems from different vendors and confirmed with reader studies, DDPET-3D consistently produced superior denoised results. Tested on images acquired at a different hospital, DDPET-3D also demonstrated its potential to be easily transferred to another medical center without further network fine-tuning. In previous literature, about 77\% of deep-learning-based healthcare papers are limited to institutional or regionally distributed dataset \cite{yang_machine_2022}. Cross-center generalizability would be crucial to implement deep-learning algorithms in clinical practice.

Compared to previous diffusion models which are time-consuming to reconstruct the entire 3D volumes, once the DDPET-3D is fully trained, the denoising process is efficient and can be finished within 15 min using one NVIDIA A40 GPU.

Compared to most of the previously published PET denoising methods \cite{zhou_federated_2023, zhou2023fedftn, xu_200x_2017, zhou2020supervised, ouyang_ultra-low-dose_2019, zhou2021mdpet, zhou2023fast, liu2022personalized, gong_pet_2019, onishi_anatomical-guided_2021, liu_pet_2022}, which learn the mapping from images collected at a specific dose level to the normal dose counterparts, the proposed DDPET-3D can be viewed as a major extension that learns to achieve dose-aware denoising by training the network using dataset with a wide range of input count levels and associating injected dose within the network. Since the noise level in PET images can be affected by various factors in clinical settings, the ability to generalize to different noise levels would be beneficial to implement the algorithm for routine clinical use.

Confirmed with reader studies by professional nuclear physicians, the proposed DDPET-3D can provide a similar or better image quality compared to normal dose images in terms of overall subjective visual impression. DDPET-3D effectively suppressed background noise and false hot spots in certain patients, making true lesions/hot spots easier to identify.

Due to the benefits mentioned above, DDPET-3D can be easily integrated with current clinical reconstruction workflow to denoise PET images.

However, this study has several limitations. \textbf{First}, as a general comparative study, DDPET-3D has not been optimized for a particular body region or patient population. Due to the high variance of tracer uptake in whole-body PET studies, training the network separately for different body regions will help improve denoising performance. \textbf{Second}, in the cross-center generalizability experiment, DDPET-3D has not been optimized for the specific hospital. In collaboration with a hospital, the proposed DDPET-3D can be further optimized with more in-house patient data. \textbf{Third}, we were unable to obtain a comprehensive clinical conclusion due to the lack of a clinical gold standard in terms of biopsy or clinical outcome. In the future, further collaboration with different medical centers would be necessary to evaluate clinical performance. We may also quantitatively measure lesion detectability using artificially inserted lesions. \textbf{Fourth}, the lesion detectability analysis was conducted using only a single subject across different noise levels, resulting in a very small sample size. A more robust evaluation would require hundreds or even thousands of lesion representations per noise level, which is currently an ongoing work. Furthermore, future studies should incorporate a wider range of lesion sizes, shapes, and anatomical locations to better reflect clinical variability. The detectability index used here serves as a simplified approximation; the use of more advanced model observers, such as the Channelized Hotelling Observer, could enhance the reliability of the results. \textbf{Lastly}, in this study, DDPET-3D was only evaluated on $^{18}\text{F-FDG}$. Further evaluations on different PET tracers would be helpful to implement the algorithms on various clinical applications. Despite these limitations, our overall results have been encouraging, indicating that DDPET-3D may offer clinical potential by achieving performance comparable to or better than normal-dose reconstructions in terms of visual quality, while also highlighting several areas for future optimization and validation.

In conclusion, the proposed DDPET-3D provides better or similar visual image quality in different low-dose settings compared to images acquired in a normal-dose setting as measured by an expert visual preference study. The experimental results showed that DDPET-3D achieved superior performance compared to previous methods. Although this study focuses on PET image denoising, we believe that the proposed technical innovations could easily be extended to other medical imaging problems, as well as to video processing and general computer vision problems involving volumetric or sequential data. In particular, the importance of maintaining slice consistency in 3D medical imaging problems is analogous to maintaining temporal consistency in video problems.

\section*{Acknowledgment}
This work was supported by the National Institute of Health (NIH) under grant numbers R01EB025468 and R01CA224140. Parts of the data used in the preparation of this article were obtained from the University of Bern, Department of Nuclear Medicine and School of Medicine, Ruijin Hospital. As such, the investigators contributed to the design and implementation of DATA and/or provided data but did not participate in the analysis or writing of this report. A complete listing of investigators can be found at: “https://ultra-low-dose-pet.grand-challenge.org/Description/”

Ramsey D. Badawi, Benjamin A. Spencer, and Reimund Bayerlein declared that UC Davis has a revenue sharing agreement with United Imaging Healthcare.

Other authors declare that they have no known conflicts of interest in terms of competing financial interests or personal relationships that could have an influence or are relevant to the work reported in this paper.

\bibliographystyle{ieeetr}
\bibliography{tmi.bib}

\end{document}